\documentclass[12pt]{article}

\usepackage{epsfig,amsmath,draftcopy}

\newcommand{\mupi}{\mu_\pi^2}
\newcommand{\mug}{\mu_G^2}
\newcommand{\rd}{\rho_D^3}
\newcommand{\rls}{\rho_{LS}^3}
\newcommand{\as}{\alpha_s}

\newcommand{\GeV}{\,\mbox{GeV}}

\newcommand\lsim{\mathop{\mbox{\vbox{\hbox{$<$} \vskip -9pt \hbox{$\sim$}
             \vskip -3pt  }}}}

\def \be{\begin{equation}}
\def \ee{\end{equation}}
\newcommand{\bea}{\begin{eqnarray}}
\newcommand{\eea}{\end{eqnarray}}
\def \nn{\nonumber}

\begin{document}
\begin{titlepage}

\vskip 8cm
\centerline{\LARGE\bf\boldmath $B$ semileptonic moments at NNLO
}

\vskip 3cm

\begin{center}
{\bf 
  Paolo Gambino \\[4mm]
\it   Dipartimento di  Fisica Teorica, Universit\`a di Torino\\[2mm] 
\& INFN, Sezione di  Torino, I-10125 Torino, Italy
}
\end{center}

\vskip 3cm

\begin{abstract}

The calculation of the moments in inclusive
$B$ meson semileptonic decays is upgraded to $O(\as^2)$.  The first three moments of the
lepton energy and invariant hadronic mass distributions are computed for arbitrary cuts on 
the lepton energy and in various renormalization 
schemes, finding in general small deviations from the $O(\as^2\beta_0)$ calculation. I also 
review the relation between $\overline{\rm MS}$ and kinetic heavy quark masses.

 
\end{abstract}

\end{titlepage}


\section{Introduction}

The  $V_{cb}$  element of the Cabibbo-Kobayashi-Maskawa quark mixing 
matrix is naturally determined in semileptonic $B$ decays to charmed hadrons. 
In the case of exclusive final states, like for $B\to D^{(*)}\ell\nu$, the corresponding
form factors have to be computed by non-perturbative methods, for instance on the lattice.
On the other hand, in the case of inclusive semileptonic decays $B\to X_c \ell\nu$
the existence of an Operator Product Expansion (OPE) ensures that  non-perturbative 
effects are suppressed by powers of the bottom mass $m_b$ and are 
parameterized by a limited number 
of matrix elements of local operators which  can be extracted from experimental data.
The total inclusive width and the first few moments of the  kinematic distributions are 
therefore expected to be well approximated by a
double series in $\as$ and $\Lambda_{\rm QCD}/m_b$ \cite{Bigi:1992su,Blok:1993va}.
The general strategy for the inclusive determination of $|V_{cb}|$ consists in extracting the 
most important non-perturbative parameters, including the heavy quark masses, from the 
moments and to employ them in the OPE expression for the total width. A determination of 
$|V_{cb}|$ follows from the comparison with the experimental total rate.

The main ingredients for an accurate analysis of the experimental data have been known for 
some time. Two implementations are currently employed by the Heavy Flavour Averaging Group (HFAG) \cite{HFAG}, based on either the kinetic scheme
\cite{kinetic,Benson:2003kp,btoc,Benson:2004sg} (see also \cite{BF} for an earlier fit) or the $1S$ scheme 
\cite{Bauer:2004ve}. They both include terms through $O(\alpha_s^2 \beta_0)$ \cite{Aquila:2005hq} 
and $O(1/m_b^3)$ \cite{1mb3} but they use different perturbative schemes and 
approximations, include 
a slightly different choice of experimental data, and estimate 
the theoretical uncertainty in two distinct ways.  According to the latest global fits 
the two methods yield very close results for $|V_{cb}|$ \cite{ckm10}.

The reliability of the inclusive method rests on our ability to control 
the higher order contributions in the double series and to constrain quark-hadron
duality violation, {\it i.e.} effects beyond the 
OPE.  The calculation of higher 
order effects allows us to verify the convergence of the double series and to
reduce and properly estimate the residual theoretical uncertainty. Duality violation effects 
\cite{Bigi:2001ys}
can be constrained {\it a posteriori}, by looking at whether 
the OPE predictions fit the experimental data. This in turn essentially depends on precise 
measurements and precise OPE predictions. 
As the experimental accuracy reached at the $B$ factories  is already better than the 
theoretical accuracy for all the measured 
moments, any effort to improve the latter is strongly motivated.

There has been recent 
progress in this direction. First,  the complete two-loop perturbative corrections to the width and to 
the moments of  the lepton energy and hadronic mass distributions have been  computed in Refs.~\cite{melnikov,czarnecki-pak,melnikov2}. This represents an important improvement
with respect to the $O(\as^2\beta_0)$ or BLM corrections, which in $B$ decays generally 
dominate the $O(\as^2)$ effects when $\as$ is normalized at $m_b$.
The main goal of this paper is to incorporate the new {\it non-BLM} corrections 
in the calculation of the semileptonic moments and to discuss their numerical impact in 
different schemes.

Higher order power corrections have also been recently considered:
a first analysis of $O(1/m_b^4)$
 and $O(1/m_Q^5)$ effects has been presented in \cite{Mannel:2010wj}.
In the higher orders of the OPE there is a proliferation of operators and therefore of non-perturbative  parameters: 
as many as nine new expectation values appear at $O(1/m_b^4)$. Since they cannot be fitted 
from experiment, the authors of  Ref.~\cite{Mannel:2010wj}
estimated them in the ground state saturation approximation and found a relatively small
+0.4\% effect on $|V_{cb}|$.
While this sets the scale of higher order power corrections, it is for the moment unclear how 
much the result depends on the use of that approximation. 

Another important source of  theoretical uncertainty are the $O(\as \Lambda^2_{\rm QCD}/
m_b^2)$ corrections to the width and to the moments. Only the $O(\as \mu^2_{\pi}/m_b^2)$
terms are known \cite{Becher:2007tk} at present. A complete 
calculation of these effects has been recently performed in the case of inclusive radiative decays \cite{ewerth}, where the $O(\as)$ correction increase the 
coefficient of $\mug$ in the rate by almost 20\%.
The extension of this calculation to the semileptonic case is in progress. 


Inclusive semileptonic $B$ decays are not only useful for the extraction of 
$V_{cb}$: the  moments are sensitive to the values of the heavy quark 
masses and in particular to a linear  combination of  $m_c$ and $m_b$ \cite{voloshin}, which to good approximation is the one needed for the extraction of $|V_{cb}|$ \cite{ckm10}.
The $b$ quark mass and the OPE expectation values obtained from the moments 
are crucial inputs in the determination of $|V_{ub}|$ from inclusive semileptonic decays, see 
{\it e.g.}\ \cite{vub} and refs.\ therein. The heavy quark masses and the OPE parameters 
are also relevant for a precise calculation of other inclusive decays like $B\to X_s 
\gamma$ \cite{gg}. On the other hand, 
$m_{c}$ and $m_b$ can now  be measured  precisely from data on charm and 
bottom production in $e^+ e^-$ annihilation  \cite{masses1,hoang}, and from 
the moments of heavy quark current correlators computed on the lattice \cite{masseslat}. 
After checking the consistency of the constraints on $m_{c,b}$ from 
semileptonic moments with these precise determinations,  
we should therefore include them as external inputs in the semileptonic fits (see \cite{ckm10} for a first 
attempt), eventually improving the accuracy of the $|V_{ub}|$ and $|V_{cb}|$ determinations.

In this paper we set up the tools for such improved analyses, extending the code
developed in \cite{btoc} and employed by HFAG to allow for NNLO fits  in arbitrary mass
schemes. In particular, in order to avoid the non-negligible uncertainty in the conversion from 
the kinetic scheme to $\overline{\rm MS}$ masses, we directly employ an $\overline{\rm MS}
$ definition for the charm quark mass. In addition, 
the study of scheme and scale dependence gives  us useful elements for an evaluation of 
the residual theoretical uncertainty. 

It is worth recalling that in the current semileptonic fits the sensitivity to $m_b$ and $\mupi$ 
is enhanced by the inclusion of the first two moments of the photon energy distribution in 
$B\to X_s \gamma$,  which are now measured with good precision. 
Their impact on the fits is 
equivalent  to that of a loose constraint on $m_b$ with 80MeV uncertainty.
However, the lower cut on the photon energy introduces a sensitivity to the Fermi motion of 
the $b$-quark inside the $B$ meson \cite{Benson:2004sg} and there are poorly known 
subdominant contributions not described by the OPE even in the absence of a photon 
energy cut \cite{paz}. 
All this makes radiative moments  different from 
semileptonic ones: in the following we will concentrate on the latter.

The paper is organized as follows: in the next Section we discuss the perturbative  
corrections of the lepton energy moments and the implementation of the NNLO contributions 
and we give tables of results in a few typical cases. In Section 3 we do the same for the 
moments of the invariant hadronic mass distribution. In Section 4 we 
review the relation between kinetic 
and $\overline{\rm MS}$ heavy quark definitions
and provide numerical  conversion formulas with uncertainty.  
Section 5 briefly summarizes the main results of the paper, while the Appendix
updates the prediction of the total rate of $B\to X_c \ell\nu$ and gives 
an approximate formula for the extraction of $|V_{cb}|$.

\section{Lepton energy moments at NNLO}
In this section we consider the first few moments of the charged lepton energy spectrum in inclusive  $b\to c\ell \nu$ decays. They are experimentally measured with high precision
(better than 0.2\% in the case of the first moment) 
but at the $B$-factories a lower cut on the lepton energy, $E_\ell \ge E_{cut}$, is applied to suppress the 
background. In fact, experiments measure  the moments at different values of $E_{cut}$. 
The relevant quantities are therefore
\begin{equation}
\langle E_\ell^{n}\rangle_{E_\ell > E_{cut}} = \frac{\int_{E_{cut}}^{E_{max}} d E_\ell \ E_\ell^n \ \frac{d\Gamma}{d E_\ell}}
{\int_{E_{cut}}^{E_{max}} d E_\ell \ \frac{d\Gamma}{d E_\ell}}\ ,
\end{equation}
for $ n $ up to 3, as well as the ratio $R^*$ between the rate with and without a cut
\begin{equation}
R^* = \label{eq:Rstar}
\frac{\int_{ E_{cut}}^{E_{max}} d E_\ell \ \frac{d\Gamma}{d E_\ell}}
{\int_0^{E_{max}} d E_\ell \ \frac{d\Gamma}{d E_\ell}}\ ,
\end{equation}
which is needed to relate the actual measurement of the rate with a cut to the total rate, from which one conventionally extracts $V_{cb}$. 
Since the physical information that can be extracted from the first three linear moments 
is highly correlated, it is convenient to study the central moments, namely the variance and asymmetry of the lepton energy distribution. In the following we will consider only $R^*$ and 
\be
\ell_1=\langle E_\ell \rangle_{E_\ell > E_{cut}}, \quad\quad \quad
\ell_{2,3}=\langle \left(E_\ell - \langle E_\ell \rangle \right)^{2,3} \rangle_{E_\ell > E_{cut}}\,.
\ee
These four observables are all functions of $m_b$ and of the two dimensionless quantities
\be
r=\frac{m_c}{m_b}, \quad \quad\quad \xi=\frac{2E_{cut}}{m_b}\ .
\ee

Our calculation follows closely the one described in \cite{btoc}. Apart from the inclusion of 
the complete $O(\as^2)$ corrections there are a few changes and improvements worth 
mentioning:  $i)$ the complete $O(\as)$  and $O(\as^2 \beta_0)$ corrections to the charged 
leptonic spectrum have been first calculated in  \cite{jez} and \cite{gremm}. While they 
can be computed numerically for any value of $\xi$ and $r$,
a numerical integration would slow down the fitting routines significantly, hence the 
need  for interpolation formulas  that must be accurate in a wide range of $\xi$ and $r$ 
values, keeping in mind that the values of $m_{c,b}$ may differ considerably from one scheme to 
another.
 The accuracy of the interpolation formulas is important because of the cancellations
that will be discussed shortly and  has  been improved wrt \cite{btoc} using 
high precision numerical results  based on \cite{Aquila:2005hq}. 
The approximations used for
the $O(\as)$  and $O(\as^2 \beta_0)$ corrections to the linear moments 
are now quite precise
and their range extends  to $0<\xi<0.75$ and $0.18< r<0.28$;
$ii)$ the code 
can now be used in the kinetic scheme at arbitrary values of the infrared cutoff $\mu$ and 
with other definition of the quark masses and of the expectation values. In particular the $\overline{\rm MS}$ definition of $m_c$ 
is now implemented;
$iii)$ in the kinetic scheme we now normalize
the matrix element of the Darwin operator, $\rd$,  at the same scale $\mu$ as the 
quark masses and the other expectation values 
(in \cite{btoc} it was normalized at $\mu=0$);
$iv)$ the code now computes the ratio $R^*$, defined in (\ref{eq:Rstar})  
and necessary to extrapolate  the rate measured at the B factories to the total
semileptonic width.
 Most of these changes  have already been included in the version of the Fortran code 
employed by HFAG \cite{HFAG} in the last few years.

The  $O(\as^2)$ corrections that are not enhanced by $\beta_0$ --- we will  
call them {\it  non-BLM corrections} --- are known to be subdominant when $\as$ is normalized at 
$m_b$. They have been 
 recently computed in \cite{melnikov,melnikov2,czarnecki-pak}. While Refs.~\cite{melnikov,melnikov2} adopt numerical methods and can take into account arbitrary cuts on 
 the lepton energy, the authors of \cite{czarnecki-pak} expand the moments in powers of 
 $m_c/m_b$ and provide only results without cuts. The two calculations are in good 
 agreement and their implementation in our codes is in principle straightforward. 
However, the strong cancellations occurring in the calculation of normalized central 
moments require a high level of numerical precision. 
Indeed, radiative corrections to the $E_l$ spectrum tend to renormalize the tree level 
spectrum in a nearly constant way, i.e.\  hard gluon emission is comparatively suppressed.
This implies that the perturbative corrections tend to drop out of
normalized moments. Let us consider for instance the first leptonic moment 
 in the kinetic scheme  with $\mu=1\GeV$, using $r=0.25$, $m_b=4.6\GeV$ and $E_{cut}=1\GeV$:
\bea
\langle E_l \rangle_{E_l>1 {\rm GeV}}& =&   1.54\GeV \left[1 +(
0.96_{den}
-0.93) \, \frac{\as}{\pi} +
(0.48_{den}
-0.46) \,\beta_0\left(\frac{\as}{\pi}\right)^2 \right.\\&&\left.
+ \left[
1.69(7)-1.75(9)_{den} 
\right] \left(\frac{\as}{\pi}\right)^2+O(1/m_b^2,\as^3)\right]\nonumber
\label{eq:cancel}\eea
It is interesting to note that such {\it kinematic} cancellations between numerator and denominator (identified by the subscript $den$) affect the  $O(\as)$,  $O(\as^2 \beta_0)$, and two-loop non-BLM corrections in  a similar way.
We have indicated in brackets the numerical uncertainty of the non-BLM correction
\cite{melnikov}: the resulting coefficient in that case is $-0.06\pm 0.12$.
Similar conclusions can be drawn at different values of the cut and for higher linear 
moments. As discussed in \cite{melnikov2}, these cancellations are not accidental. 
 In the limit  $\xi\to \xi_{max}=1-r^2$
the cancellations between numerator and denominator are complete at any perturbative
order: therefore the  higher the cut,  the  stronger the cancellation. Moreover the
peak of the lepton energy distribution is  relatively narrow and close to the endpoint, which 
further protects the moments from radiative corrections.

In the case of the higher central moments, additional cancellations occur at each 
perturbative order between normalized moments. In $\ell_2$, for instance, $\langle E_l^2 \rangle$ and $\langle E_l \rangle^2$ tend to cancel each other: for the same inputs as in Eq.(\ref{eq:cancel}) we have 
\be
\ell_2= \langle E_\ell^2\rangle -\langle E_\ell\rangle^2=(2.479-2.393)\GeV^2= 0.087\GeV^2.\nonumber
\ee
Such cancellations are quite 
general  and are further enhanced by higher $E_{cut}$. They are simply a 
consequence of the fact that, as we have just seen, at each perturbative order the spectrum 
follows approximately the tree-level spectrum, which is peaked at $\xi\approx 0.7-0.8$.

One obvious consequence of the cancellations we have just discussed is that the numerical 
accuracy with which the non-BLM corrections are known becomes an issue. While the origin of the cancellations is understood,  we need a precise calculation to know 
their exact extent, and the result will have some impact on the estimate of the remaining 
theoretical uncertainty. 

The building blocks of the perturbative calculation are the adimensional moments 
\be
L_n  = \frac1{\Gamma_0}  \frac1{m_b^{n}} {\int_{ E_{cut}}^{E_{max}} d E_\ell \ E_\ell^n \ \frac{d\Gamma}{d E_\ell}}
\end{equation}
where $\Gamma_0$ is the total width at the tree-level. Any $L_n$ can be expanded in $\as$ and $1/m_b$
\be
L_{n}= L_{n}^{(0)} + \frac{\as(m_b)}{\pi} L_{n}^{(1)} + \left(\frac{\as}{\pi}\right)^2
\left( \beta_0 \,L_{n}^{({\rm BLM})}+ L_{n}^{(2)}\right) + L_{n}^{(pow)} +...\nonumber
\ee
where  we have distinguished 
between BLM and non-BLM  two-loop corrections ($\beta_0=11-\frac23 n_l$) and indicated 
by $L_{n}^{(pow)}$ the power-suppressed contributions.
The expansion for  the normalized moments is
\be 
\langle E_\ell^n\rangle = \frac{L_{n}^{(0)}}{L_0^{(0)}} \left[ 1+ 
 \frac{\as(m_b)}{\pi} \eta_{n}^{(1)} + \left(\frac{\as}{\pi}\right)^2
\left( \beta_0 \,\eta_{n}^{({\rm BLM})}+ \eta_{n}^{(2)} - \eta_n^{(1)} \frac{L_0^{(1)}}{L_0^{(0)}} 
\right) + \eta_{n}^{(pow)} +...\right],
\ee
where 
\be
\eta_i^{(a)}= \frac{L_i^{(a)}}{L_i^{(0)}}-\frac{L_0^{(a)}}{L_0^{(0)}}\,,
\ee
and an analogous formula holds for $R^*$.

Both available non-BLM calculations have been performed in the on-shell scheme and 
give us results for $L_n^{(2)}$ at different values of $r$ with an uncertainty  due either to the 
numerical integration (and therefore of statistical origin)
or to the truncation of the $r$ expansion. However,  $L_n^{(2)}$ has been computed 
at $\xi\neq 0$ only numerically  \cite{melnikov2}.
The two calculations can be  combined in order to reduce the final uncertainty.
In the analytic calculation of \cite{czarnecki-pak} the expansion of the $O(\as^2)$ corrections 
to the moments at $\xi=0$ includes at most $O(r^7)$ terms and converges quite slowly 
for $r\sim 0.2-0.25$, in the relevant physical range. We can take the size of the last 
term included,  $O(r^7)$, as a rough estimate of its non-gaussian uncertainty. It 
turns then out that the combinations  of  two-loop non-BLM integrals $L^{(2)}_i$
that enter the normalized moments, $\eta_i^{(2)}$, 
at $\xi=0$ and $r\lsim 0.26$ can be  more accurately determined using 
\cite{czarnecki-pak} than using  the tables of \cite{melnikov2}. 
\begin{figure}[t]
\begin{center}
\includegraphics[width=8.cm]{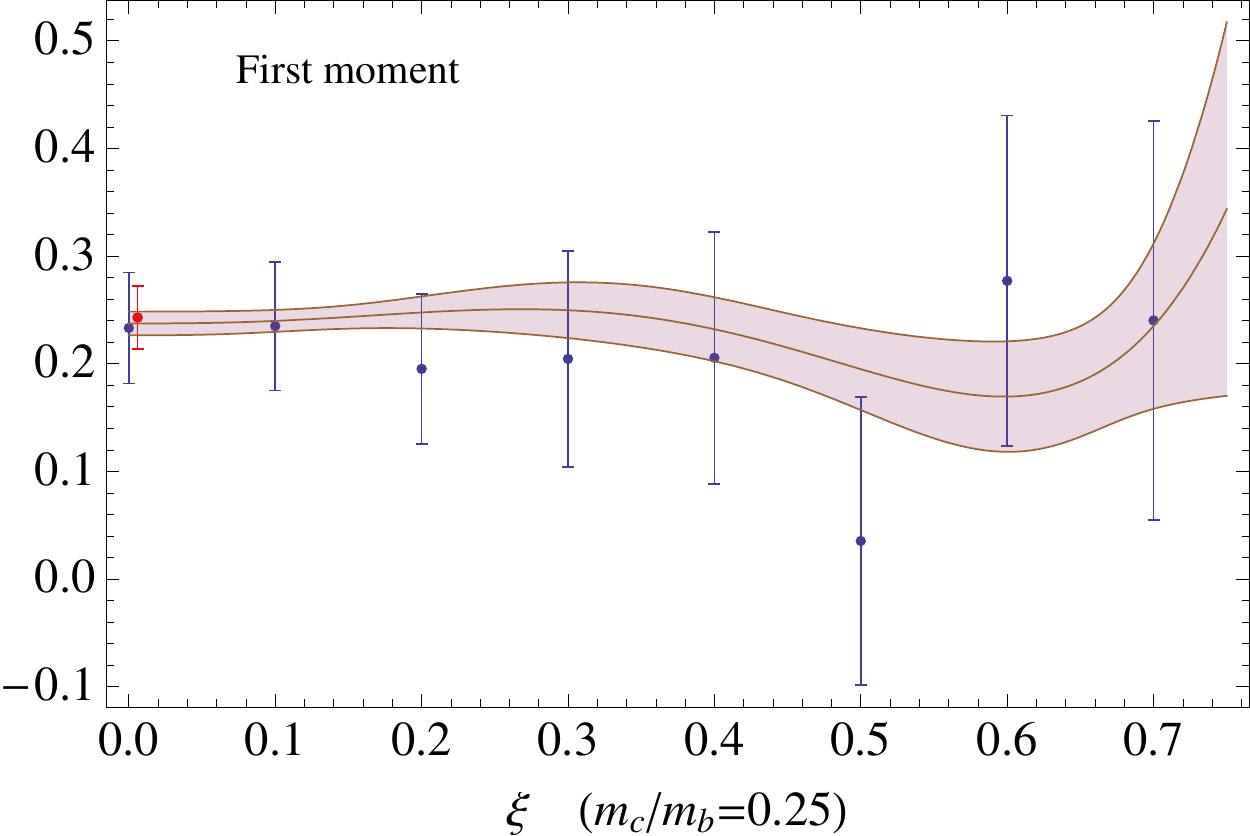}
\includegraphics[width=7.8cm]{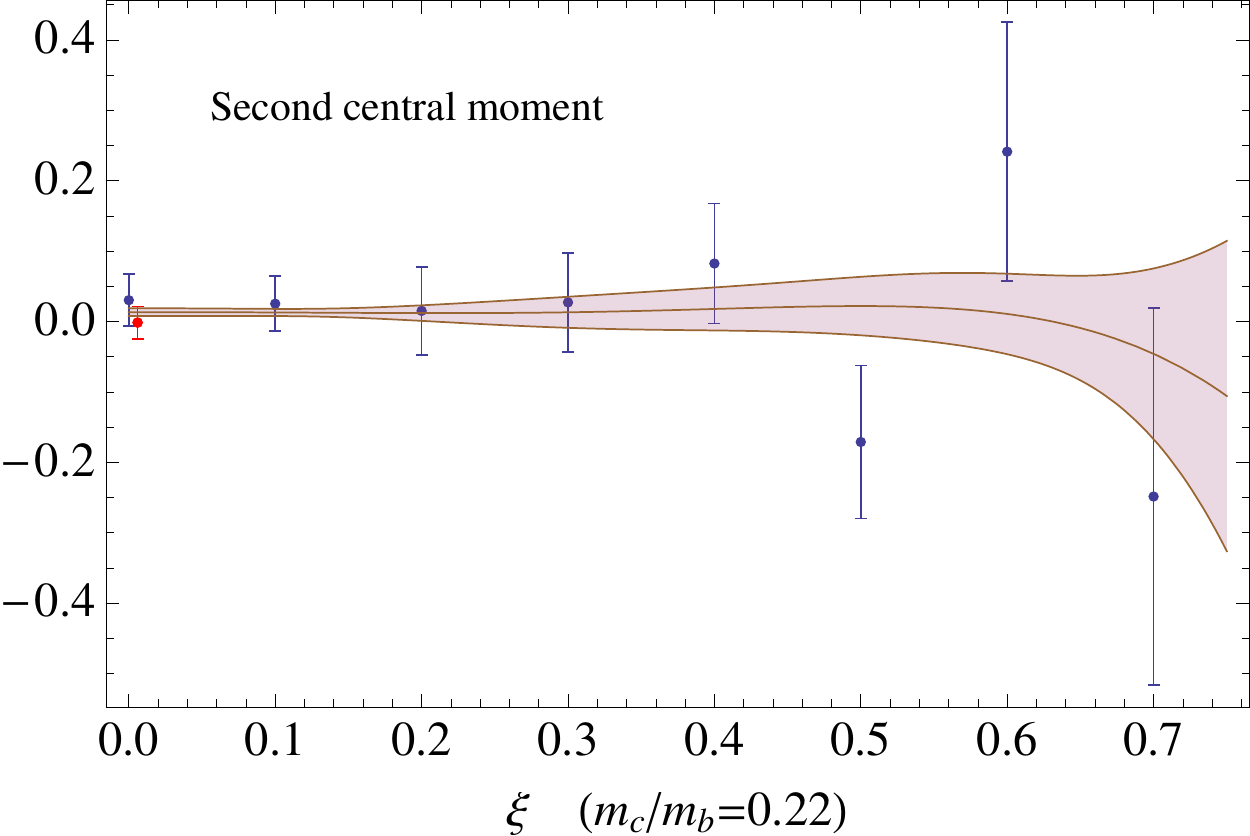}\vskip 3mm
\includegraphics[width=8cm]{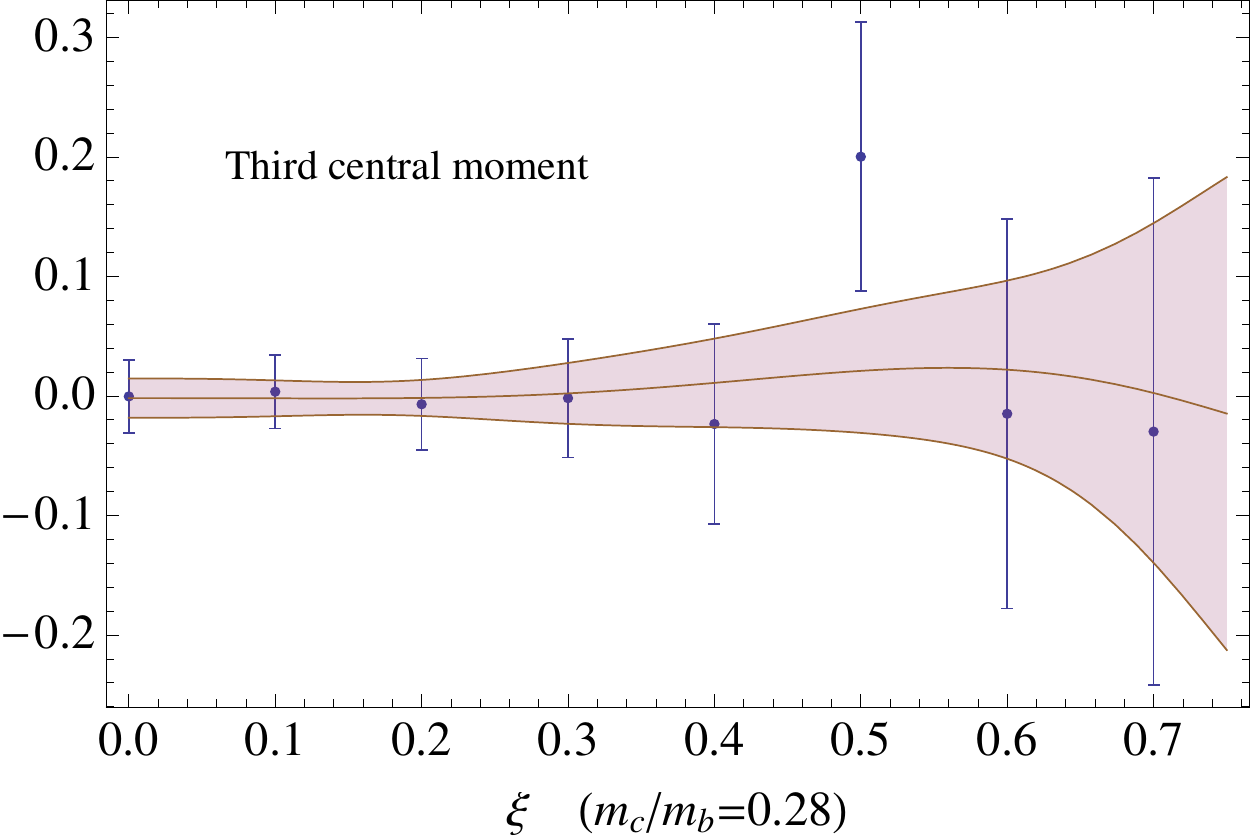}
\includegraphics[width=7.8cm]{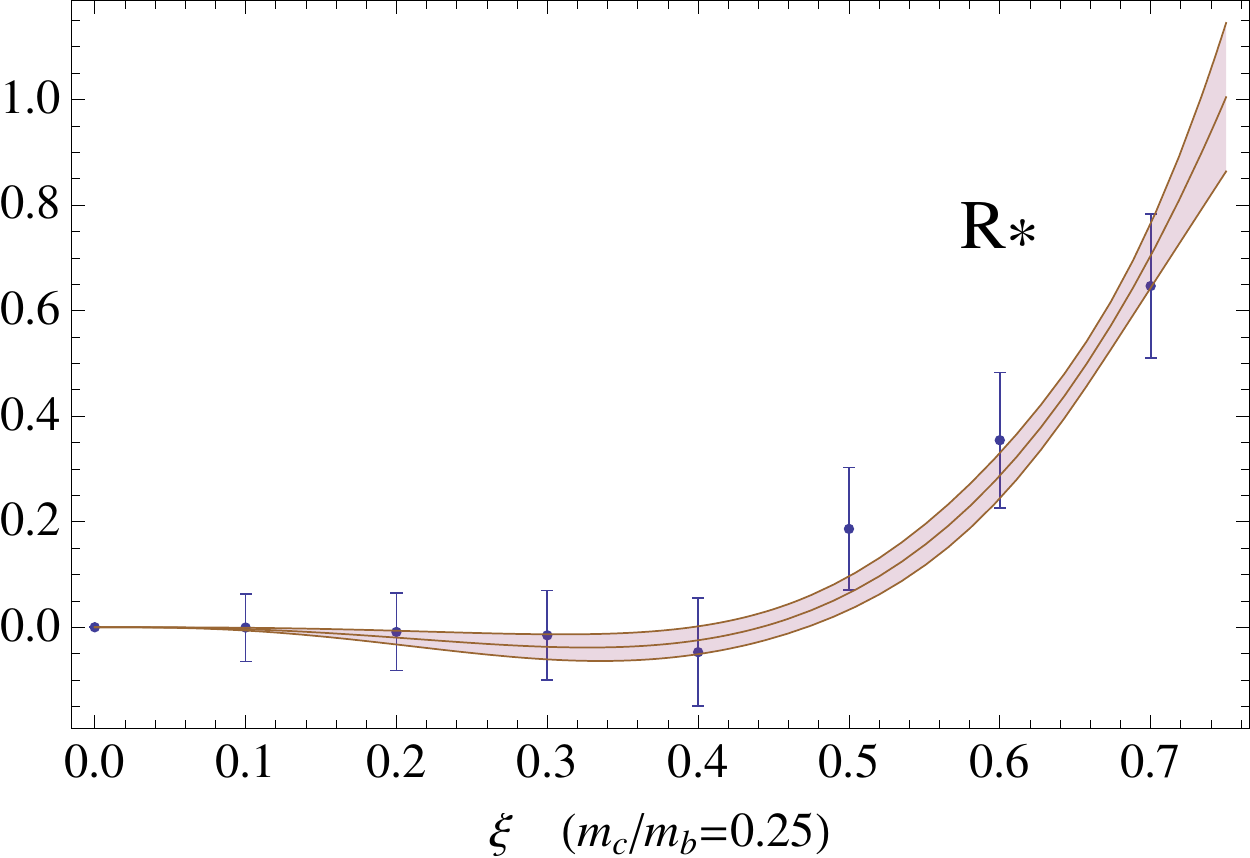}
\caption{\sf Combinations 
of two-loop non-BLM contributions 
entering the normalized central leptonic moments and $R^*$: numerical evaluation 
\cite{melnikov2} with blue errors and analytic one \cite{czarnecki-pak} at $\xi=0$ with red errors vs.\ fits (shaded bands).}
\label{fig:fits}
\end{center}
\end{figure}

 It is also helpful to notice that, since the  electron energy spectrum must vanish at low 
 energies at least like $E_\ell^2$, only terms  $O(\xi^3)$ and higher are relevant: the two-loop 
 corrections $\eta_i^{(2)}$ must be flat for small $\xi$ and the results of \cite{melnikov2} are perfectly consistent with this requirement. One can therefore perform a fit
to all the available results at different $r$ and $\xi$ values using simple functions of $\xi$ 
and $r$.   We have checked that, given 
the level of accuracy provided in \cite{melnikov2}, it is sufficient to consider an expansion up 
to $O(\xi^5)$   with coefficients  linearly  dependent on $r$. The $\chi^2/dof$ is
generally low, about 0.4, as the spread of central values in the low-$\xi$ region is
much smaller than the size of the errors. 
We also know that the functions $\eta_i^{(2)}(\xi,r)$ must vanish linearly at the 
endpoint $\xi=1-r^2$, and one can 
implement this constraint in the fit, but the result are unchanged for $\xi<0.7$.

\begin{figure}[t]
\begin{center}
\includegraphics[width=7.8cm]{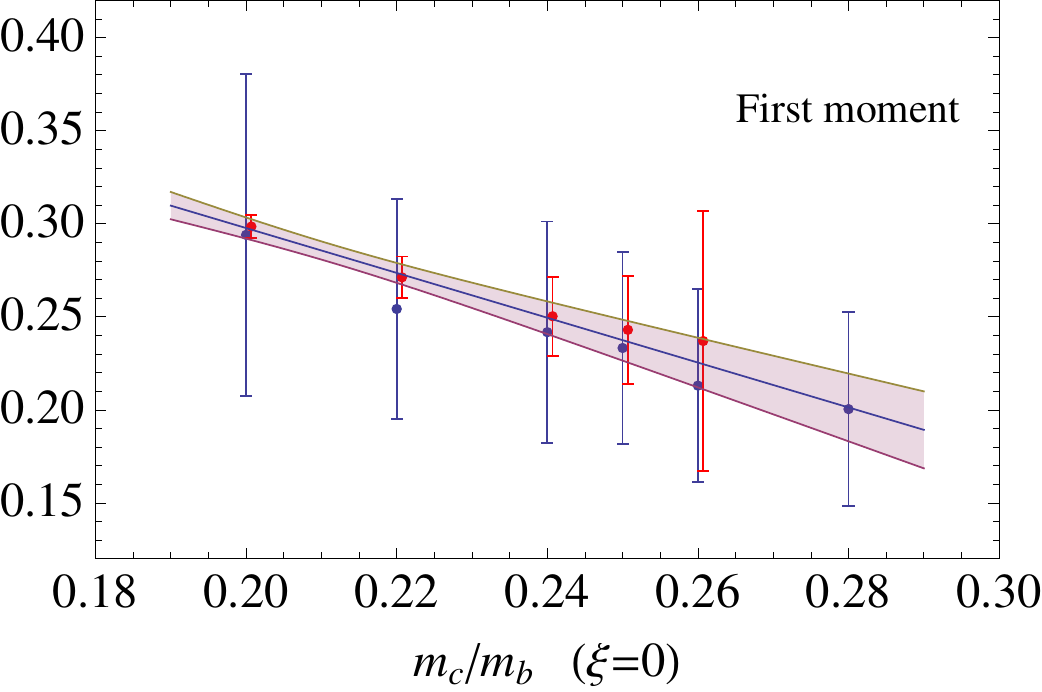}
\caption{\sf The non-BLM contribution $\eta_1^{(2)}(\xi=0)$ to the first leptonic moment as a function of $r$: numerical evaluation \cite{melnikov2} with blue errors and analytic one \cite{czarnecki-pak} at $\xi=0$ with red errors vs.\ fits (shaded bands).}
\label{fig:fits2}
\end{center}
\end{figure}

In the first plot in Fig.~\ref{fig:fits} we show the combination $\eta_1^{(2)}$  that enters the 
first  normalized leptonic moment $\ell_1$  as a function of $\xi$ for $r=0.25$. 
In the plot we compare 
the numerical evaluations given in Table 1 of  \cite{melnikov2} and their associated error 
bars with the fit. At $\xi=0$ we also show the result of \cite{czarnecki-pak}, whose
uncertainty is estimated as explained above.
The shaded band represents the 1$\sigma$ uncertainty of the fit.
In the case of Eq.~(\ref{eq:cancel}) the error of the non-BLM $O(\as^2)$ 
coefficient estimated in this way is $\pm0.03$, with a sizable reduction wrt that equation.
 For higher cuts and for values  of $r$ at the edge of the range $0.2\le r\le 0.28$ considered in 
  \cite{melnikov2} the reduction is weaker.
In Fig.~\ref{fig:fits2} we show the $r$-dependence of $\eta_1^{(2)}(\xi=0)$, comparing the fit 
with the existing results. It is clear that a linear fit is perfectly adequate.  The errors
obtained by combining in quadrature the numerical uncertainties of
$L_1^{(2)}$ and $L_0^{(2)}$ appear  overestimated.
We have found similar results for the higher linear moments, but we show in Fig.\ref{fig:fits}
only the combinations that actually enter the second and third central moments,
\bea
\eta_{2c}^{(2)}&=& w_2 \,\eta_2^{(2)} - 2\,w_1^2 \,\eta_1^{(2)} \nonumber\\
\eta_{3c}^{(2)}&=&w_3 \,\eta_3^{(2)} +3 w_1\left(2 w_1^2-w_2 \right) \eta_1^{(2)} - 3\,w_1\, w_2  \,\eta_2^{(2)}, \nonumber
\eea
where $w_i=L_i^{(0)}/L_0^{(0)}$ are tree-level functions.
Despite the residual uncertainty,  the new results confirm that the pattern of 
cancellations observed at $O(\as)$ and $O(\as^2\beta_0)$ carries on at the complete $O
(\as^2)$. 
This is illustrated in Tables \ref{tab:1} and \ref{tab:2} for  $E_{cut}=0$ and 1GeV. In 
these Tables  we report  the values of the first three central moments for the reference 
values of the input parameters
\bea\label{inputs}
m_b=4.6 \GeV, \ \ \ m_c=1.15\GeV, &&\  \mupi=0.4\GeV^2,\ \ \ \mug=0.35\GeV^2 \nonumber\\
\rd=0.2\GeV^3, &&\ \rls=-0.15\GeV^3,
\eea
 and for $\mu=0$ (corresponding to the on-shell scheme) and $\mu=1$GeV.  The 
kinetic scheme expressions are obtained from the on-shell ones after re-expressing
the pole quark masses and their analogues for $\mupi$  and $\rd$ in terms 
of low energy running quantities, employing the  two-loop 
 contributions computed in \cite{Czarnecki:1997sz} and re-expanding all perturbative series.\footnote{Unlike the $O(\as^2)$ calculation of the moments, the kinetic scheme expressions
 of Ref.~\cite{Czarnecki:1997sz} do not explicitly include finite charm mass effects in the 
 loops.
 This mismatch will be ignored in the following: it is effectively equivalent to a perturbative 
 redefinition of our mass parameters and it is relevant only when we translate them to 
 other schemes, see Section 4. \label{foot:1}}
 
 In our calculation we remove all terms of $O(\as \Lambda^2/m_b^2)$ because 
 they are not yet known completely, 
 but we retain suppressed terms of $O(\as^2 \mu^3/m_b^3)$ that originate in the kinetic 
 scheme from a simultaneous perturbative shift in $m_b$ and $\mupi$ or $\rd$, although 
 they    turn out to be completely negligible in the case of leptonic 
 moments.
  
\begin{table}
  \begin{center} \begin{tabular}{|c|lll|lll|}
    \hline 
     & &$\mu=0$ & & &$\mu=1$GeV & \\
    \hline
    & $\ell_1$ & $\ell_2$ & $\ell_3$ & $\ell_1$ & $\ell_2$ & $\ell_3$  \\ \hline
       tree & 1.4131 &0.1825& -0.0408 & 1.4131 &0.1825 & -0.0408 \\
    $1/m_b^3$& 1.3807 & 0.1808 & -0.0354 & 1.3807 & 0.1808 & -0.0354 \\
$O(\as)$ & 1.3790  & 0.1786 & -0.0354 & 1.3853 & 0.1811 &-0.0349 \\
$O(\beta_0\as^2)$ & 1.3731  & 0.1766 & -0.0350(1) & 1.3869 & 0.1820 &-0.0341(1) \\
$O(\as^2)$ & 1.3746(1)  & 0.1767(2) & -0.0349(6) & 1.3865(1) & 0.1816(2) &-0.0340(6) \\
tot error \cite{btoc} &  & & & 0.0125 & 0.0055 & \ 0.0026\\
    \hline 
  \end{tabular} \end{center}
    \caption{\sf \label{tab:1} The first three leptonic moments for the reference values of the 
    input parameters and $E_{cut}=0$, in the on-shell and kinetic schemes. In parentheses the
numerical uncertainty of the BLM and non-BLM contributions (see text).
  } 
\end{table}
\begin{table}
  \begin{center} \begin{tabular}{|c|llll|}
    \hline 
    & $\ell_1$ & $\ell_2$ & $\ell_3$ &$R^*$ \\ \hline
     &\multicolumn{4}{|c|}{$\mu=0$}  \\
    \hline
    tree & 1.5674 &0.0864 & -0.0027 & 0.8148\\
 $1/m_b^3$ &1.5426 &0.0848 & -0.0010&  0.8003 \\
$O(\as)$ & 1.5398  & 0.0835 & -0.0010 & 0.8009\\
$O(\beta_0\as^2)$ & 1.5343  & 0.0818(1) & -0.0009(2) & 0.7992\\
$O(\as^2)$ & 1.5357(2)  & 0.0821(6) & -0.0011(16) & 0.7992(1)\\
    \hline 
     & \multicolumn{4}{|c|}{$\mu=1$GeV}  \\ \hline
$O(\as)$ &  1.5455 & 0.0858 &-0.0003 &0.8029\\
$O(\beta_0\as^2)$ & 1.5468 & 0.0868(1) &\ 0.0005(2) & 0.8035 \\
$O(\as^2)$ &  1.5466(2) & 0.0866 & \ 0.0002(16)  &0.8028(1)\\
$O(\as^2)^{**}$ &  \ -- & 0.0865 &\ 0.0004 & \ --\\
tot error \cite{btoc}&    0.0113 & 0.0051 & \ 0.0022 &\\
    \hline 
  \end{tabular} \end{center}
    \caption{\sf \label{tab:2} The first three leptonic moments for the reference values of the input parameters and $E_{cut}=1$GeV, in the on-shell and kinetic schemes.  } 
\end{table}

Non-BLM contributions to the leptonic moments are generally tiny in the kinetic scheme.
This is not necessarily the case if we adopt other renormalization schemes. We illustrate the 
point by using an $\overline{\rm MS}$ definition of the charm quark mass. As mentioned in 
the Introduction, this may be useful for the inclusion in semileptonic fits of precise mass 
constraints. Table \ref{tab:3} shows the results at $E_{cut}=1\GeV$ in the case the charm 
quark mass is renormalized in the $\overline{\rm MS}$ scheme at two different values of the 
$\overline{\rm MS}$ scale; $\bar\mu=2 $ and 
3 \GeV. The bottom mass and the OPE parameters are still defined in the kinetic scheme.
We observe that in this  case the non-BLM corrections are not always negligible, especially 
at $\bar\mu=3\GeV$. This is due to the fact that the $\overline{\rm MS}$ running is 
numerically important and driven by non-BLM effects. The sizable
non-BLM corrections found in this case are therefore related to the change of scheme for 
$m_c$ which is known with high accuracy.
\begin{table}
  \begin{center} \begin{tabular}{|c|llll|}
    \hline 
     & \multicolumn{4}{|c|}{$\mu=1$GeV,\  $m_c^{\overline{\rm MS}}(2\rm GeV)$ }   \\
    \hline
    & $\ell_1$ & $\ell_2$ & $\ell_3$ & $R^*$   \\ \hline
tree &1.5792 & 0.0890 & -0.0032 & 0.8200\\
$1/m_b^3$ &1.5536 & 0.0873 & -0.0013 & 0.8058\\
$O(\as)$ &1.5502 & 0.0869 & -0.0003 &0.8056  \\
$O(\beta_0\as^2)$ &1.5540 &0.0884(1) &\ 0.0004(2) &  0.8073 \\
$O(\as^2)$ &1.5523(3) & 0.0879(6)& -0.0002(16)& 0.8061(1)  \\
$O(\as^2)^{**}$ & \ --  & 0.0878& \ 0.0004&  \ --   \\
\hline
    &     \multicolumn{4}{|c|}{$\mu=1$GeV,\  $m_c^{\overline{\rm MS}}(3\rm GeV)$ }\\
    \hline
        & $\ell_1$ & $\ell_2$ & $\ell_3$ & $R^*$   \\ \hline
tree & 1.6021 & 0.0940 & -0.0043 & 0.8296\\
$1/m_b^3$ &1.5748 & 0.0922 & -0.0020 &  0.8159\\
$O(\as)$ & 1.5613  & 0.0894 & -0.0004 & 0.8118 \\
$O(\beta_0\as^2)$ & 1.5629  & 0.0904(1) & \ 0.0004(2) & 0.8125 \\
$O(\as^2)$ & 1.5571(4)  & 0.0890(9) & -0.0008(25) & 0.8090(2)\\ 
$O(\as^2)^{**}$ &  \ -- & 0.0889 & \ 0.0006 & \ -- \\ \hline
  \end{tabular} \end{center}
    \caption{\sf \label{tab:3} The first three leptonic moments for the reference values of the 
    input parameters and $E_{cut}=1$GeV, in the kinetic scheme with $\overline{\rm MS}$ charm 
    mass evaluated at $\mu=2$ and 3GeV, with $m_c(2{\rm GeV})=1.1$GeV and $m_c(3{\rm 
    GeV})=1$GeV. The uncertainty in the $O(\as^2)$ is larger in the second case because the $m_c/m_b$ value is closer to the edge of the range considered in \cite{melnikov2}. 
      } 
\end{table}

To better understand what drives larger corrections in the $\overline{\rm MS}$ scheme,
let us now consider $\ell_{1,2,3}$ for $E_{cut}=1\GeV$ and reference inputs. A shift in $m_{b,c}$ induces in the central leptonic moments the shifts
\bea
&&\delta\ell_1= 0.325\, \delta m_b -0.237\, \delta m_c, \quad\quad
\delta\ell_2= 0.094\, \delta m_b -0.058\, \delta m_c, \nonumber \\ 
&&\hspace{3cm}\delta\ell_3= 0.009\, \delta m_b -0.010\, \delta m_c,
\label{schemechange}
\eea
which vanish for $\delta m_c/\delta m_b\approx 1.4$, 1.6 and 0.9, respectively. Therefore,
the leptonic central moments, and particularly the first two, give similar constraints in the $(m_c,m_b)$ plane. In the case of a change of mass 
scheme, the bulk of the shift in the leptonic moments is given by the above $\delta \ell_i$,
where $\delta m_{c,b}$ represent the difference between the on-shell masses and the 
masses in the new scheme.
In the kinetic scheme with  $\mu=1\GeV$ we have $\delta m_c/\delta m_b\approx 1.2$ at $O
(\as^2)$, which explains why the change of scheme does not spoil the cancellations in the 
perturbative corrections, as shown in Table 2.
In other schemes and for different values of  $\mu$ 
the situation can be different, and indeed   when we adopt 
 the $\overline{\rm 
MS}$ scheme for $m_c$ with $\bar\mu=3\GeV$ (lower sector of Table 
\ref{tab:3}) we observe slightly larger perturbative corrections. 
As our calculation is complete at NNLO, one could worry that higher orders 
in $\delta m_{c,b}$ could spoil the cancellations that occur in the on-shell scheme 
and lead to an underestimate of higher order corrections. This appears to be unlikely, except 
for large $\overline{\rm MS}$ scales; even that unnatural case, however, would be under 
control, as the $O(\as^3)$  $\bar\mu$ evolution is known.
\begin{table}
  \begin{center} \begin{tabular}{|c|llll|}
    \hline 
   $E_{cut}$ 
   & $\frac{\eta_1^{(2)}}{\eta_1^{(\rm BLM)}}$ &$\frac{\eta_{2}^{(2)}}{\eta_2^{(\rm BLM)}}$ &$\frac{\eta_3^{(2)}}{\eta_3^{(\rm BLM)}}$ &$\frac{\eta_{2c}^{(2)}}{\eta_{2c}^{(\rm BLM)}}$ \\ \hline
  0  
   &-0.28(1) & -0.26(1) &-0.26(1) &-0.15(9)\\
0.6 
&-0.32(3) & -0.29(1) &-0.28(1) &-0.14(19)\\
0.9 
&-0.33(4) & -0.32(2) & -0.30(1)&-0.21(30) \\
1.2 
&-0.30(6) & -0.32(3) &-0.30(2)&-0.36(62)\\
1.5 
&-0.33(9) & -0.31(4) &-0.27(2) &+0.3(1.4)\\
    \hline 
  \end{tabular} \end{center}
    \caption{\sf \label{tab:ratioslept} 
    Ratio of non-BLM to BLM contributions to the leptonic moments at various 
    $E_{cut}$ values, in the on-shell scheme. } 
\end{table}

While in the case of $\ell_1$ and $R^*$ the numerical accuracy is always adequate for the 
fits to experimental data, 
the poor direct knowledge of irreducible non-BLM corrections to the 
second and especially the third central moment suggests to adopt for them a different 
approach. We have already argued that the two-loop calculations  confirm that the
cancellations occurring at $O(\as)$ and $O(\as^2\beta_0)$ replicate at $O(\as^2)$ as well. 
\begin{figure}[t]
\begin{center}
\includegraphics[width=5.2cm]{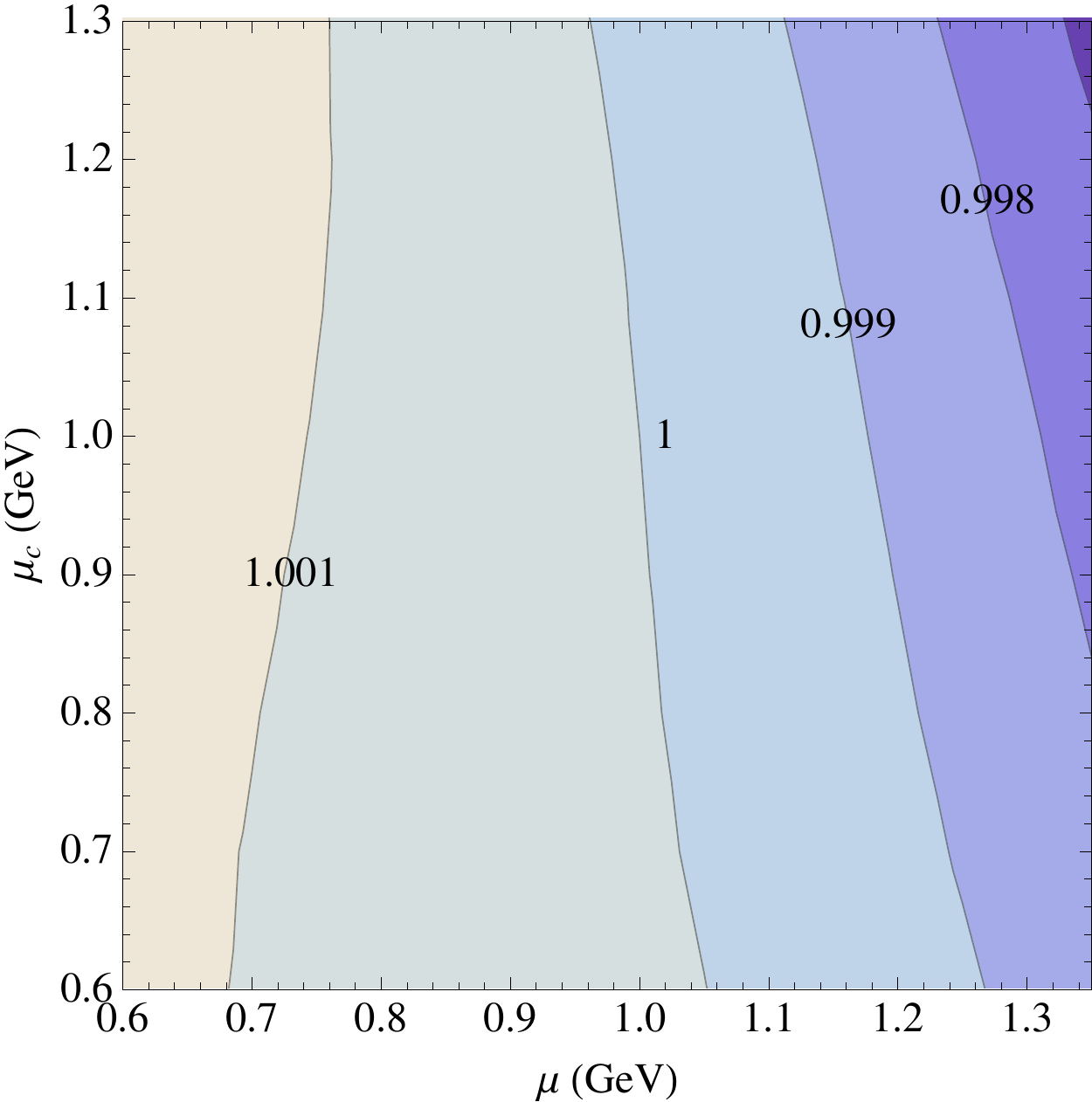}\ 
\includegraphics[width=5.2cm]{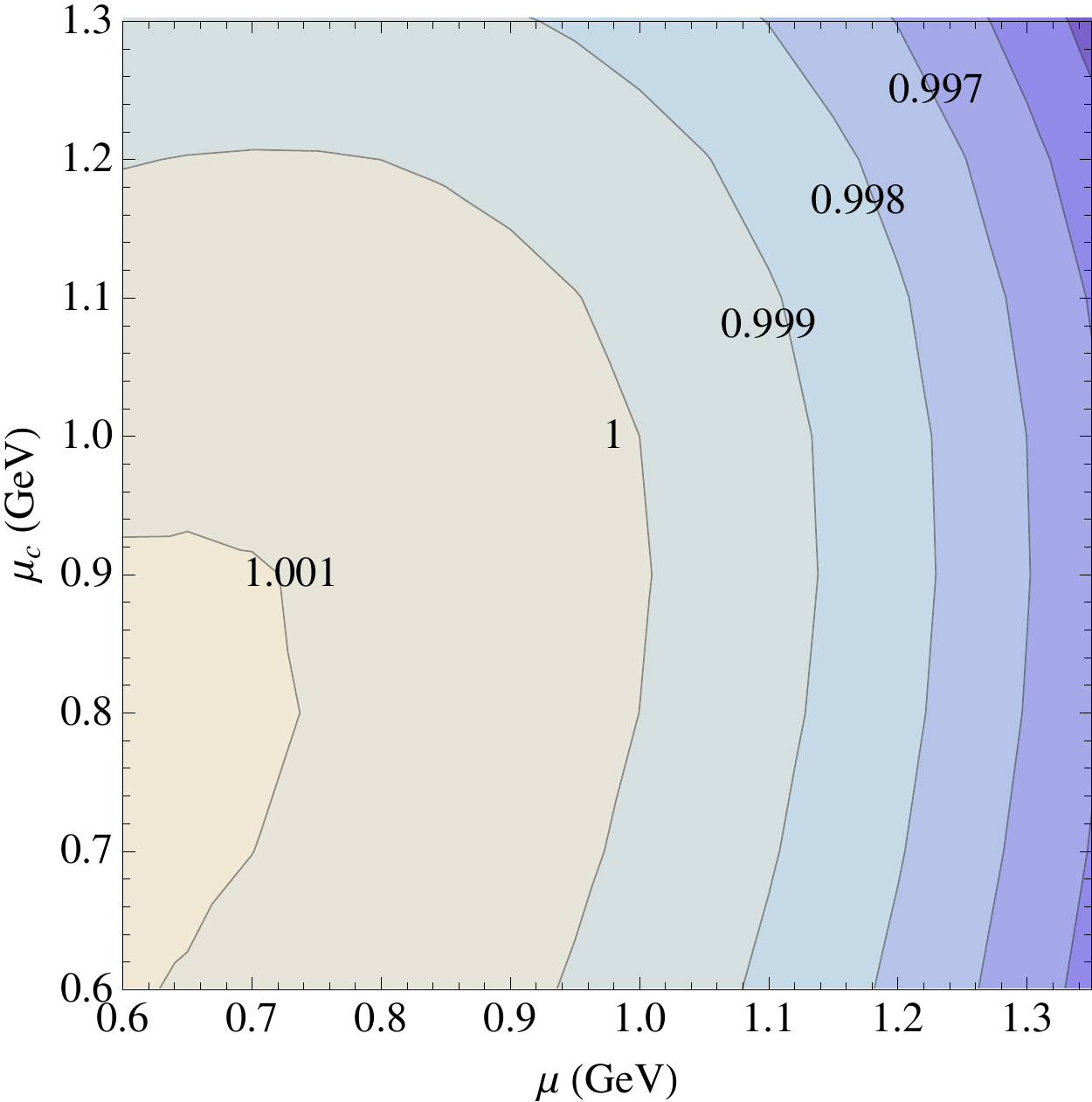}\
\includegraphics[width=5.2cm]{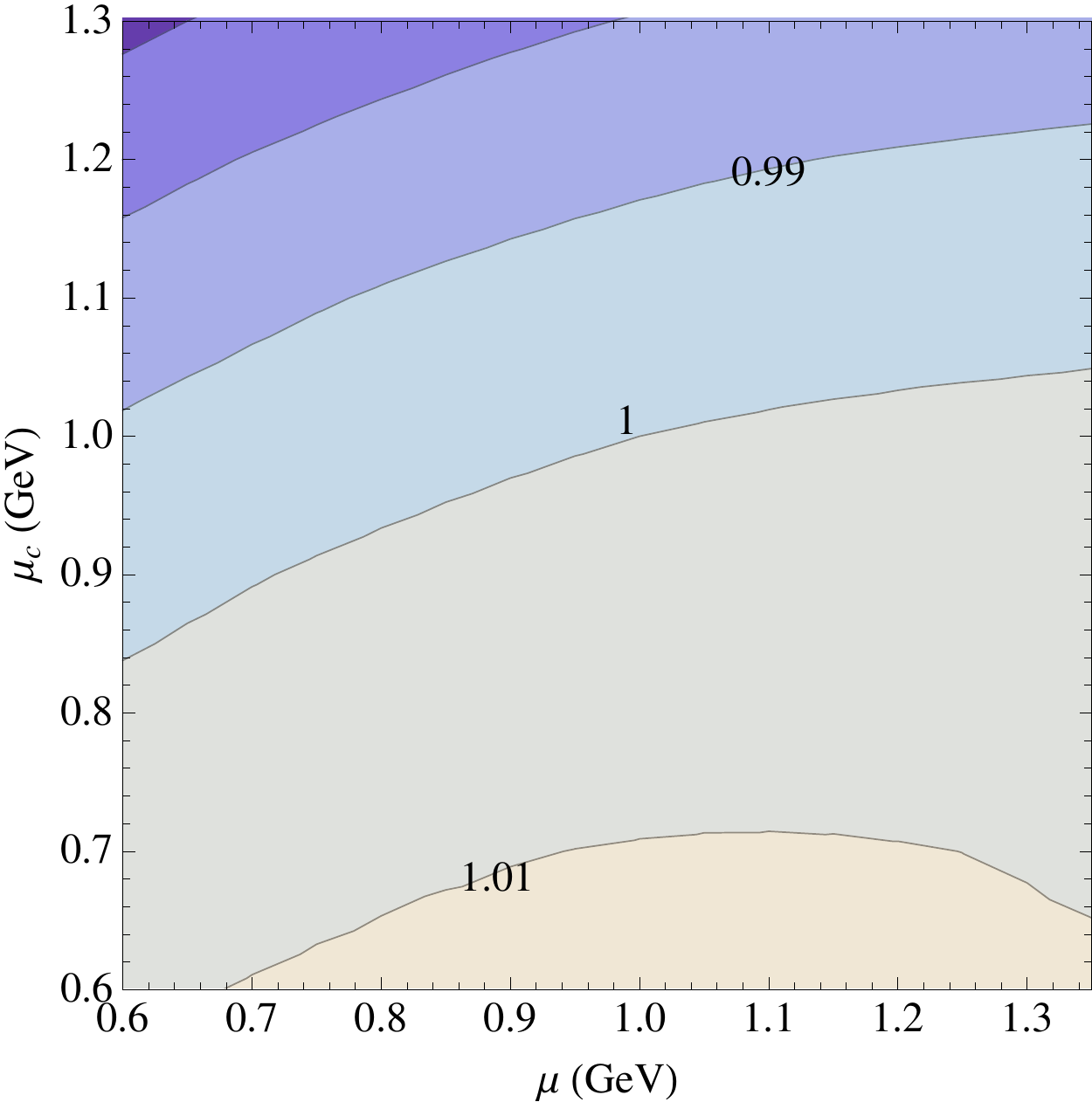} 
\caption{\sf $\mu$ and $\mu_c$ dependence of the first three leptonic central moments at $E_{cut}=0$ normalized to their reference value $\mu=\mu_c=1\GeV$. The three plots refer to $\ell_{1,2,3}$, respectively.}
\label{figkinmu}
\end{center}
\end{figure}
This is further illustrated by Table \ref{tab:ratioslept} 
where the ratio between the non-BLM and BLM contributions to $\eta_{1,2,3}$ are shown 
and, as usual,  the uncertainty is dominated by that of the non-BLM fits. The ratio of 
non-BLM to BLM contributions is well determined and close to $-0.3$, but in the case of
$\eta_{2c}^{(2)}$ the uncertainty is much larger, especially at large cuts, namely where the 
cancellations are necessarily enhanced. We therefore choose to assume a value 
$-0.2\pm0.1$ for this ratio and adopt it as default. In the case of $\eta_{3c}^{(2)}$ the 
uncertainty is so much bigger than the effect that we simply remove this contribution, 
retaining however reducible non-BLM contributions that are important in the $\overline{\rm 
MS}$ scheme. Reference values for this default choice are reported in Tables \ref{tab:2}, \ref{tab:3} in the row denoted by $**$. 

\subsection{Scale dependence of leptonic moments}
\begin{figure}[h]
\begin{center}
\includegraphics[width=7cm]{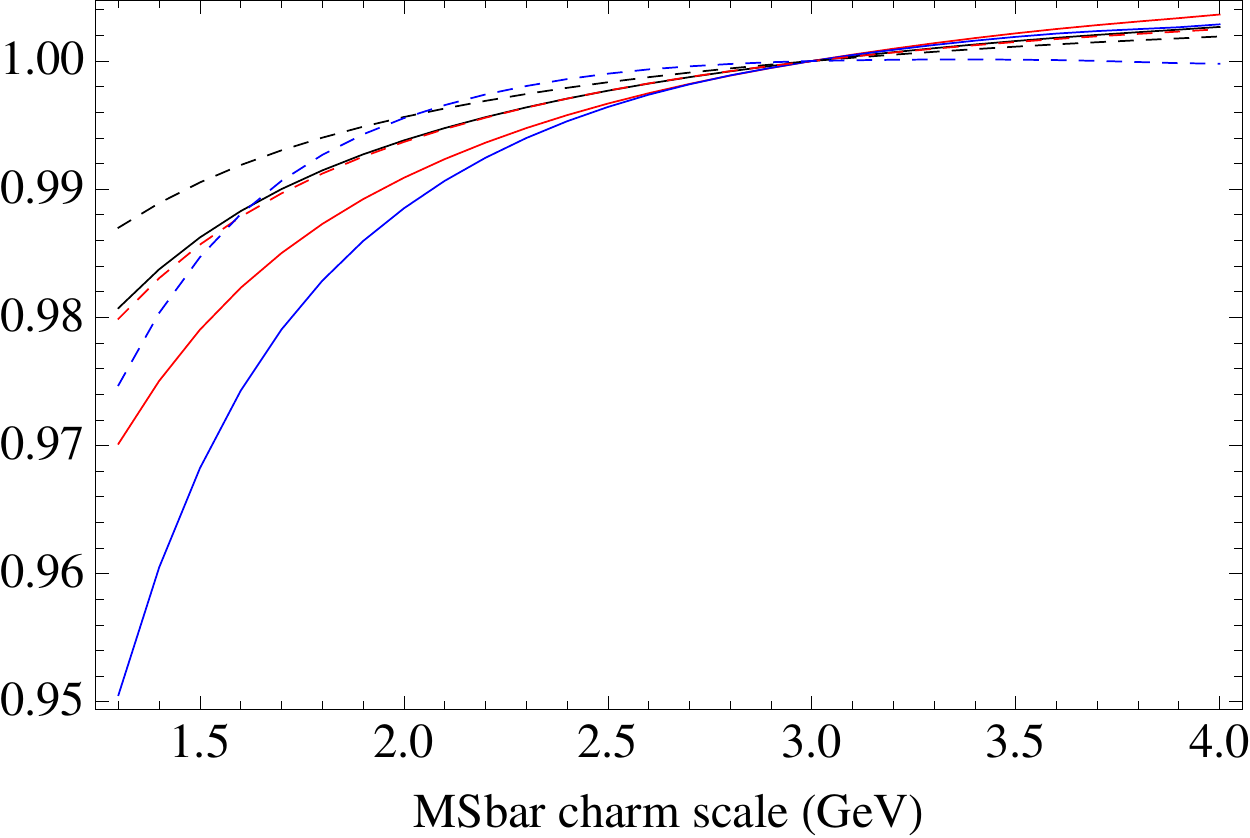} 
\includegraphics[width=7cm]{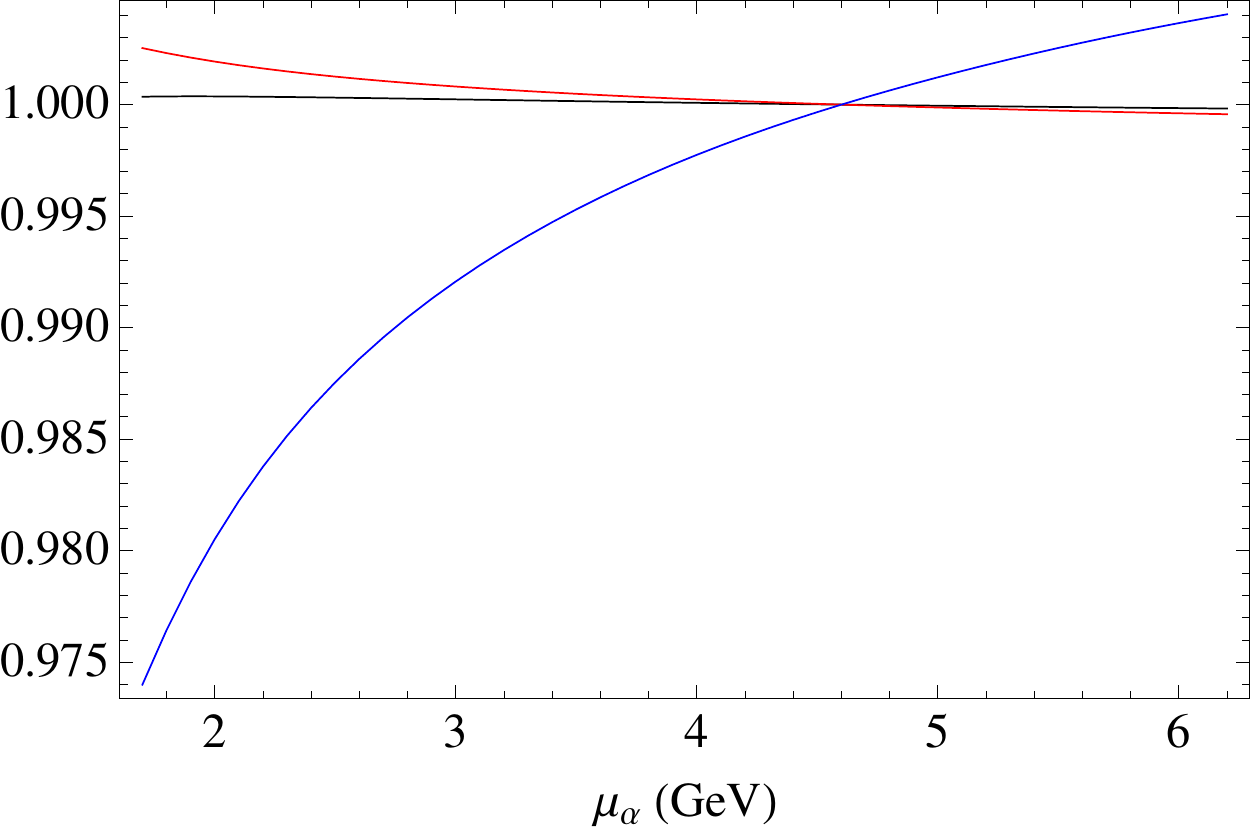} 
\caption{\sf Dependence of the first three leptonic central moments at $E_{cut}=0$ on 
$a)$ the charm $\overline{\rm MS}$ scale $\bar\mu$ (left panel) and $b)$  the scale of $\as$ 
when the kinetic scheme is applied to $m_c$ as well (right panel). The moments are 
normalized to their values at $\bar\mu=3\GeV$ and $\mu_\alpha=m_b$, respectively.
 The kinetic cutoff is fixed at 1 \GeV. The black, red, 
 blue lines refer to $\ell_{1,2,3}$. The solid (dashed) lines refer to four (two) loop evolution.}
\label{figmsmu1}
\end{center}
\end{figure}

The size of uncalculated higher order perturbative corrections can be estimated in various 
ways. We have already made a few remarks in this direction and we have studied 
the size of the NNLO contributions in various mass schemes. Now we study how  our 
predictions for the leptonic moments depend on various unphysical 
scales that enter the calculation. 
In Fig.~\ref{figkinmu} we show the dependence of the leptonic moments in the kinetic 
scheme on the cutoff $\mu$ and on $\mu_c$, the cutoff related to the kinetic 
definition of the charm 
mass. Indeed, there is no reason of principle to set $\mu_c=\mu$ as we have done above.
In the plots we have taken the $\mu=\mu_c=1\GeV$ values of the $b$ and $c$ masses and 
of the OPE parameters $\mupi$ and $\rd$ in 
Eq.~(\ref{inputs}) and have evolved them to different $\mu,\mu_c$ using $O(\as^2)$ 
perturbative expansions \cite{Czarnecki:1997sz} with $\as(m_b)$. 
The plots refer to 
$E_{cut}=0$, but the absolute size of the scale dependence is similar at different cuts, while 
the relative size varies in an obvious way.
Notice that for values of $\mu$ significantly higher than 1\GeV\ 
the kinetic definition may  lead to artificially large 
perturbative corrections, especially  in the case of the charm mass. In the acceptable range
shown in the plots the scale dependence is small, and only for $\ell_1$ it is larger than 
the whole $O(\as^2)$ correction.

  The plot on the left of Fig.~\ref{figmsmu1} refers instead to the calculation with the charm mass  defined in 
  the $\overline{\rm MS}$. It shows the
dependence on  the charm mass scale $\bar \mu$ for fixed kinetic cutoff $\mu=1\GeV$, 
normalized to the values $\bar\mu=3\GeV$, $m_c(3\GeV)=1\GeV$. We have evolved $\overline{ m}_c$ using the four loop Renormalization Group Evolution (RGE) implemented in \cite{rundec}.
It is clear that in this case higher order corrections present in the four loop RGE (and to a lesser extent even in the two loop RGE, see Fig.~\ref{figmsmu1})
lead to very sizable corrections and that scales below $2\GeV$ should  be 
avoided. 
Even above $2\GeV$ the observed scale dependence of $\ell_1$ is larger than its whole 
perturbative contribution.
This is  not an indication of large higher order corrections in the calculation
of the moments, but only in the $\overline{\rm MS}$ evolution of $m_c$. The best procedure
to extract $m_c$ is therefore to compute moments at a scale $\bar\mu\approx 2\GeV$ that 
tends to minimize the corrections to the moments,  and then to evolve to other scales using 
the most accurate RGE formula.

Finally, in the right-hand plot of Fig.~\ref{figmsmu1} we show the dependence of the kinetic 
scheme results on the $ \overline{\rm MS}$ scale of $\as$, $\mu_\alpha$, for fixed cutoff $
\mu=\mu_c=1\GeV$. The evolution of $\as$ is 
again computed at 4 loops, using  $\as(m_b)=0.22$ as input. Due to its direct connection 
with the size of the perturbative contributions,  the $\mu_\alpha$ dependence  is generally 
tiny, with the only exception of $\ell_3$. 
For instance,  using $\as(m_b/2)$ instead of $\as(m_b)$ in Table 1,  we would have $\ell_1= 1.3870\GeV$ in place of 1.3865\GeV.  

We will briefly come back to the scheme dependence of the leptonic moments in Sec.~4. 
Before turning to the hadronic moments, however, it is useful to spend a few words on the 
estimate of the total theoretical uncertainty of the leptonic moments, a subject of great 
importance for the global fits from which we extract $|V_{cb}|$.
In Ref.~\cite{btoc} and in the fits based on it, the overall theoretical uncertainty related to 
$\ell_i$ was computed 
by adding in quadrature various contributions: the shits in $\ell_i$ due to a $\pm20\%$ ($
\pm30\%$) variation in $\mupi$ and $\mug$ ($\rd$ and $\rls$), a $\pm20$MeV variation 
in $m_{c,b}$, and a $\pm 0.04$ variation in $\as$.  The uncertainty estimated in this way
is reported in the last row of Tables \ref{tab:1} and \ref{tab:2}: in all cases it is about 
five times  larger than  the current experimental one.
Following the inclusion of the 
NNLO calculation we can now slightly reduce this estimate.
Keeping in mind  the remaining sources of theoretical uncertainty\footnote{While 
the $O(\as\mu_\pi^2/m_b^2)$ contributions to $\ell_i$  are tiny \cite{Becher:2007tk}, the $O(1/m_b^{4,5})$ corrections estimated in Ref.~\cite{Mannel:2010wj} are almost as large as  the total error in the last row of Tables \ref{tab:1} and \ref{tab:2}.}, one could use the same 
method with a $\pm10$ MeV variation in  $m_{c,b}$  and $\pm 0.02$ in $\as$; in this way 
the total theoretical uncertainty in $\ell_1$ is reduced by $\sim 25\%$ wrt the last row in 
Tables \ref{tab:1} and \ref{tab:2}, while the improvement is only minor for $\ell_{2,3}$.

\section{Hadronic mass moments at NNLO}
At the time Ref.~\cite{btoc} was   published not even all the 
$O(\as)$ corrections to the hadronic moments were available for  $\xi\neq 0$, and 
therefore the original calculation in the kinetic scheme was rather incomplete.
 The original code underwent a subsequent upgrade to include the full $O(\beta_0 \as^2)$, 
 computed numerically following  \cite{Aquila:2005hq} and implemented through 
 precise interpolation formulas,
 and in this form has been employed in recent HFAG  analyses.
Here we discuss the implementation of the complete NNLO corrections and present a few 
results. The code for the hadronic moments shares the same features of the leptonic 
code, including  flexibility in the choice of the scheme and of all the relevant scales.

We recall that due to bound state effects the invariant mass $M_X$  of the hadronic system,
which is measured experimentally,
is related to the invariant mass $m_x$ and energy $e_x$ of the partonic (quarks and gluons) 
final state by
\be
M_X^2= m_x^2+ 2 \,e_x \,\overline{\Lambda} +\overline{\Lambda}^2,
\label{eq:MX}
\ee
where $\overline{\Lambda}=M_B-m_b$. Since $\overline{\Lambda}\approx 0.7$GeV is 
relatively large, we do not expand in $\overline{\Lambda}/m_b$ and retain all powers of
$\overline{\Lambda}$. It follows from Eq.~(\ref{eq:MX}) that the $M_X^2$ moments
are linear combinations of the moments of $m_x^2$ and $e_x$, namely of the building blocks
 \be
M_{ij}= 
\frac1{\Gamma_0} \frac1{m_b^{2i+j}}\int_{E_\ell>E_{cut}} (m_x^{2}-m_c^2)^i \, e_x^{j}\,d \Gamma \, .
\ee
For instance, in the case of the first normalized moment we have
\be
\langle M_X^2\rangle= m_c^2+ 
\overline{\Lambda}^2+m_b^2 \,\frac{M_{10}}{M_{00}}+ 2 \,m_b\, \overline{\Lambda} \,\frac{M_{01}}{M_{00}}. \nonumber
\ee
 Like in the case of leptonic moments, we consider only central 
 higher moments, and specifically
\be
h_1=\langle M_X^2\rangle, \quad \quad
h_2=\langle (M_X^2-\langle M_X^2\rangle)^{2}\rangle,\quad\quad
h_3=\langle (M_X^2-\langle M_X^2\rangle)^{3}\rangle \, .
\ee
 The calculation of $h_{1,2,3}$ requires the knowledge of the building blocks $M_{ij}$ with $i+j\le 1,2,3$, respectively. They receive both 
perturbative and non-perturbative power-suppressed contributions:
\be
M_{ij}= M_{ij}^{(0)} + \frac{\as(m_b)}{\pi} M_{ij}^{(1)} + \left(\frac{\as}{\pi}\right)^2
\left( \beta_0 M_{ij}^{({\rm BLM})}+ M_{ij}^{(2)}\right) + M_{ij}^{(pow)} +...
\ee
where the tree-level contributions $M_{ij}^{(0)}$ vanish for $i>0$ and we have distinguished 
between BLM and non-BLM  two-loop corrections ($\beta_0=11-\frac23 n_l$, $n_l=3$).
Since we are interested in the {\it normalized} moments, 
we will be mostly concerned with the combinations
\be
\chi_{ij}^{(a)}=\frac{M_{ij}^{(a)}}{M_{00}^{(0)}}- \frac{M_{00}^{(a)}M_{ij}^{(0)}}{(M_{00}^{(0)})^2},
\ee
with $a=1,{\rm BLM},2, pow$, which enter the perturbative and non-perturbative expansion for the normalized moments and are functions of $\xi$ and $r$:
\be
\frac{M_{ij}}{M_{00}}=\frac{M_{ij}^{(0)}}{M_{00}^{(0)}} +
\frac{\as(m_b)}{\pi} \chi_{ij}^{(1)} + \left(\frac{\as}{\pi}\right)^2
\left( \beta_0 \chi_{ij}^{({\rm BLM})}+ \chi_{ij}^{(2)}-\frac{M_{00}^{(1)}}{M_{00}^{(0)}} \chi_{ij}^{(1)} \right) + \chi_{ij}^{(pow)} +...
\ee
 \begin{figure}[t]
\begin{center}
\includegraphics[width=8.cm]{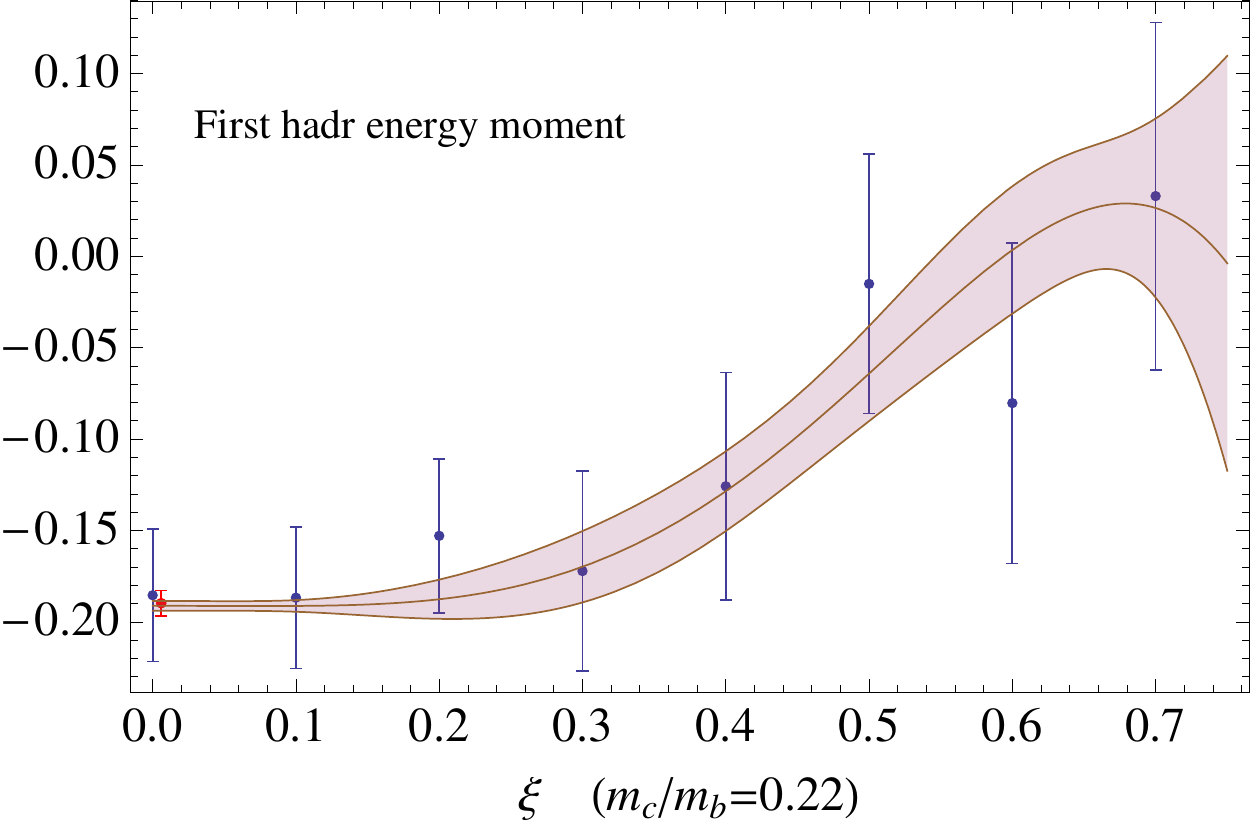}
\includegraphics[width=8cm]{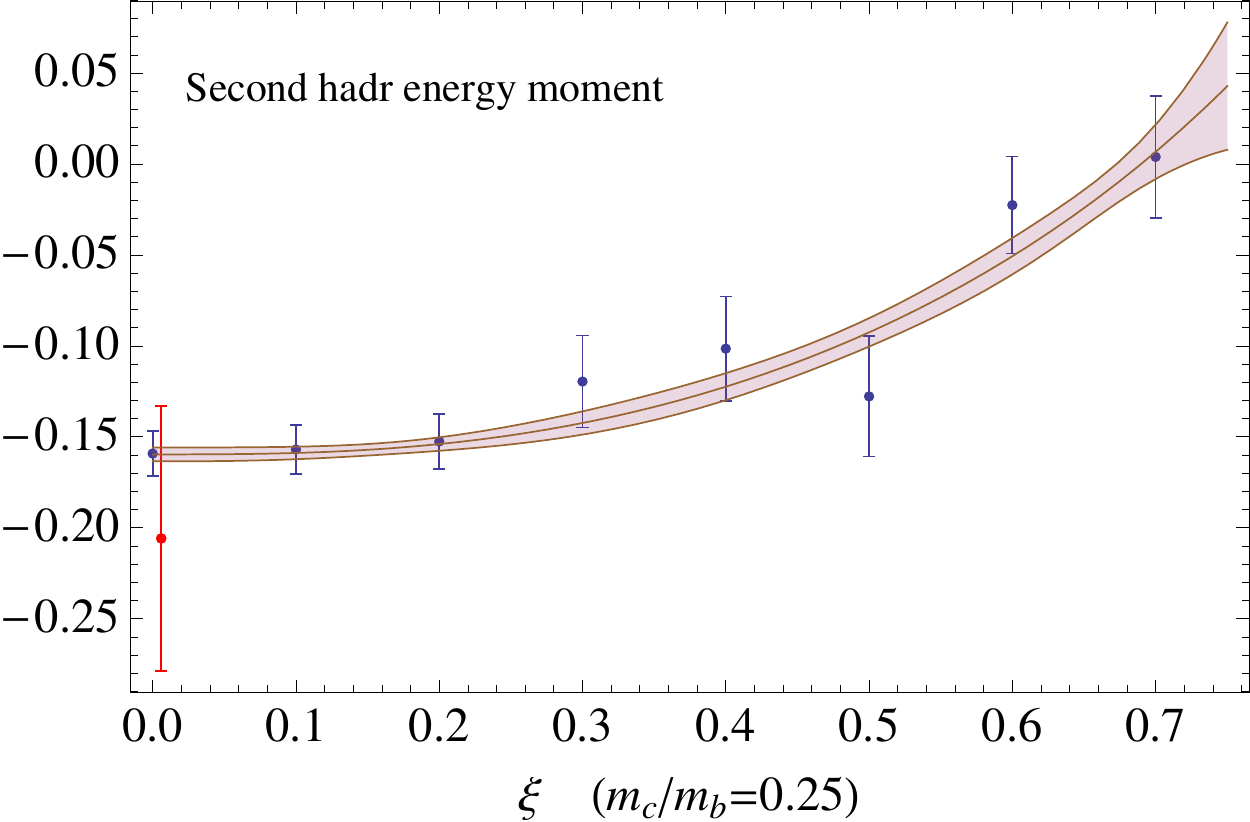} \vskip 3mm
\includegraphics[width=8cm]{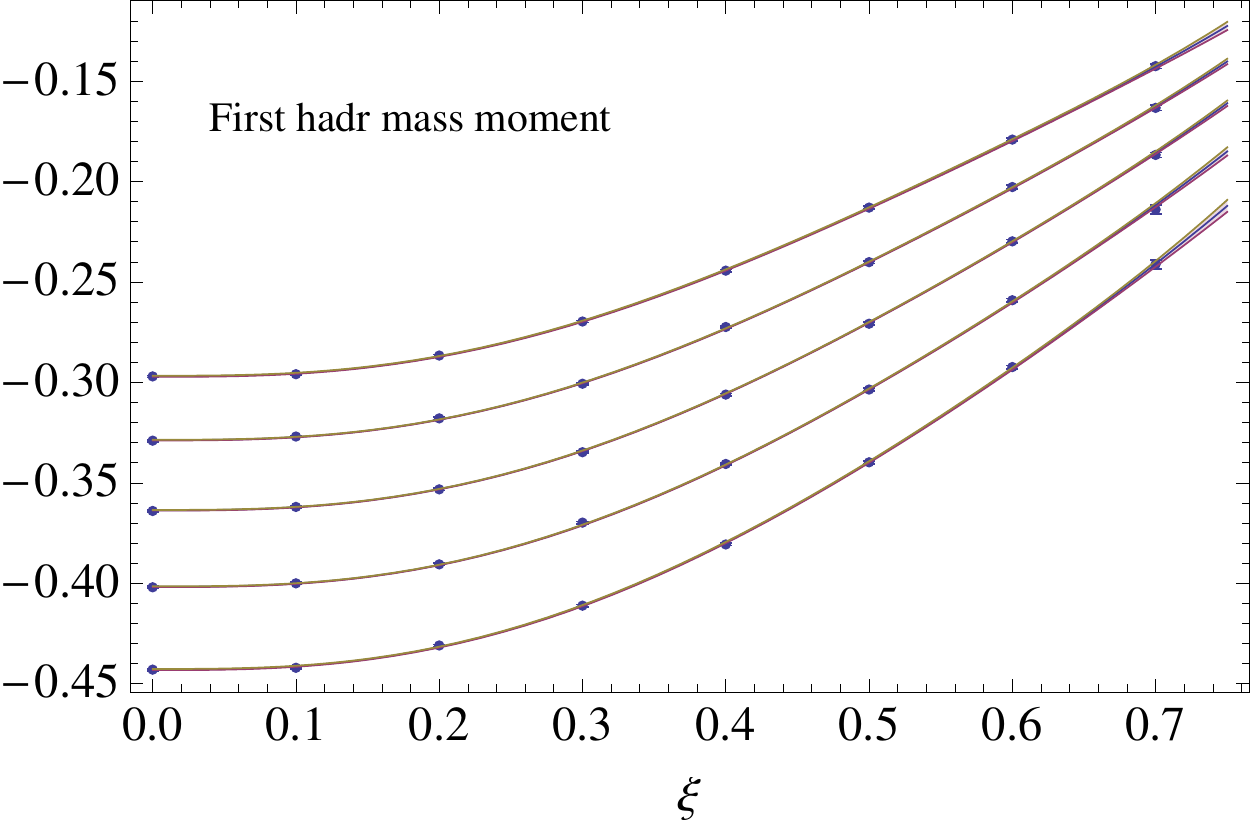}
\caption{\sf Combinations 
of two-loop non-BLM contributions $\chi_{01}^{(2)}$,  $\chi_{02}^{(2)}$, 
and $\chi_{10}^{(2)}$ 
entering the hadronic moments: numerical evaluation 
\cite{melnikov2} with blue errors and analytic one \cite{czarnecki-pak} at $\xi=0$ with red errors vs.\ fits (shaded bands). $\chi_{10}^{(2)}$ (lower panel) is shown for $r=0.2,0.22,0.24,0.26,0.28$.}
\label{fig:hadfits}
\end{center}
\end{figure}

The first $O(\as)$ calculation appeared in  \cite{Falk:1997jq} and was later completed 
in \cite{trott,kolyapert,Aquila:2005hq}, 
while the  $O(\as^2\beta_0)$ (BLM) calculation was presented in
\cite{kolyapert,Aquila:2005hq}. Analytic expressions for the $O(\as)$
contributions to the building blocks $M_{ij}$ with $i\neq 0$ are
available as functions of $\xi,r$ \cite{Aquila:2005hq}. For the
remaining building blocks and for all the $O(\as^2\beta_0)$ we employ
high-precision two-parameters interpolation formulas to the results of the numerical code of 
Ref.~\cite{Aquila:2005hq}, which are  valid in the range $0<\xi<0.8$ and $0.18<r<0.29$. 
The use of these approximations induces a negligible error.

 The complete NNLO results presented here are based on the on-shell scheme results of
Refs.~\cite{czarnecki-pak,melnikov2}. Ref.~\cite{czarnecki-pak} provides us with 
NNLO contributions to the first two $e_x$ moments at $\xi=0$, and we have the non-BLM 
contributions to 
several of the building blocks from Ref.~\cite{melnikov2}. In all the available cases we 
proceed in the same way as for leptonic moments and perform fits to the tables of 
\cite{melnikov2} and to the $\xi=0$ of \cite{czarnecki-pak}; some of the results are shown in 
Fig.~\ref{fig:hadfits}. Unfortunately not all  the building 
blocks needed for the second and third $M_X^2$ moments are reported in the tables of \cite{melnikov2}: 
the missing building blocks are $M_{20}^{(2)}$, $M_{21}^{(2)}$, and $M_{30}^{(2)}$, which 
have been computed only for  $r=0.25$ and for $\xi=0$ and 0.435 \cite{melnikov-private}.

In order to deal with these three cases we observe that in the on-shell scheme the ratio of 
non-BLM to BLM contributions to $\chi_{ij}$ is always remarkably insensitive to the value of 
$\xi$, as shown in Table \ref{tab:ratioshadr}, and in fact much more insensitive than the  
analogous ratios to the tree-level or one-loop contributions.
 \begin{table}
  \begin{center} \begin{tabular}{|c|lllll|}
    \hline 
   $E_{cut}$ &
      $\frac{\chi_{01}^{(2)}}{\chi_{01}^{(2,\rm BLM)}}$ & $\frac{\chi_{10}^{(2)}}{\chi_{10}^{(2,\rm BLM)} }$  & $\frac{\chi_{02}^{(2)}}{\chi_{02}^{(2,\rm BLM)} }$ & $\frac{\chi_{11}^{(2)}}{\chi_{11}^{(2,\rm BLM)} }$& $\frac{\chi_{12}^{(2)}}{\chi_{12}^{(2,\rm BLM)} }$ \\ \hline
  0  & 
  -0.24(1)& -0.24 & -0.23(1)&-0.24 &-0.24\\
0.6 & 
  -0.25(2) & -0.24& -0.24(1)&-0.24&-0.24\\
0.9 & 
 -0.26(3) & -0.23& -0.24(1)&-0.23&-0.24\\
1.2 & 
  -0.26(5)& -0.23& -0.24(2)&-0.23&-0.23\\
1.5 &  
-0.14(20)&-0.23& -0.17(7)&-0.23&-0.24\\
    \hline 
  \end{tabular} \end{center}
    \caption{\sf \label{tab:ratioshadr} 
    Ratio of various perturbative contributions relevant to the  hadronic moments at various 
    $E_{cut}$ values, in the on-shell scheme. 
  } 
\end{table}
In particular, this ratio is known precisely for $i\neq 0$ and varies by about $\pm 0.01$
in the range $0.2<r < 0.28$. In view of the other sources of theoretical errors, it is  therefore 
sufficient to estimate the missing $\chi_{20}^{(2)}, \chi_{30}^{(2)},\chi_{21}^{(2)}$ using the 
values of their ratios in the two available points, and to assign them  a $\pm 0.02$ error:
\be
\frac{\chi_{20}^{(2)}}{\chi_{20}^{(\rm BLM)}}=-0.215\pm 0.020, \quad\quad
\frac{\chi_{21}^{(2)}}{\chi_{21}^{(\rm BLM)}}=-0.215\pm 0.020, \quad \quad
\frac{\chi_{30}^{(2)}}{\chi_{30}^{(\rm BLM)}}=-0.205\pm 0.020
\ee

\begin{table}
  \begin{center} \begin{tabular}{|c|lll|lll|}
    \hline 
     &   \multicolumn{3}{|c|}{$\mu=0$} &    \multicolumn{3}{|c|}{$\mu=1$GeV } \\
    \hline
    & $h_1$ & $h_2$ & $h_3$ & $h_1$ & $h_2$ & $h_3$  \\ \hline
LO &     4.420 & 0.199  & -0.03&4.420 &0.199 & -0.03\\
power & 4.515 &0.767  &5.51  &4.515  &0.767  & 5.51 \\
$O(\as)$ & 4.662  & 1.239 & 7.78 & 4.504 &  1.069&6.13 \\
$O(\beta_0\as^2)$ & 4.834  &1.661  & 9.51 &4.493 &1.301  &6.39 \\
$O(\as^2)$ & 4.811  & 1.603(5) & 9.22(7) & 4.495 & 1.222(5) &6.24(7) \\
tot error \cite{btoc}&  & & & 0.143 &0.510  &1.38 \\
    \hline 
  \end{tabular} \end{center}
    \caption{\sf \label{tab:6} The first three hadronic moments for the reference values of the input parameters and $E_{cut}=0$, in the on-shell and kinetic schemes. 
  } 
\end{table}\begin{table}
  \begin{center} \begin{tabular}{|c|lll|lll|}
    \hline 
     &        \multicolumn{3}{|c|}{$\mu=0$} &    \multicolumn{3}{|c|}{$\mu=1$GeV } \\
    \hline
    & $h_1$ & $h_2$ & $h_3$ & $h_1$ & $h_2$ & $h_3$  \\ \hline
    LO &     4.345 &0.198   & -0.02&4.345 & 0.198&-0.02 \\
 $1/m_b^3$ & 4.452 & 0.515 &  4.90&4.452  &0.515  &4.90  \\
$O(\as)$ & 4.563  & 0.814 & 5.96 & 4.426 &  0.723& 4.50\\
$O(\beta_0\as^2)$ & 4.701  & 1.105 &6.85 &4.404  & 0.894 & 4.08\\
$O(\as^2)$ & 4.682(1)  & 1.066(3)  &6.69(4)  &4.411(1)  &0.832(4)  & 4.08(4)\\
tot error \cite{btoc}&  & & &  0.149& 0.501 & 1.20\\
    \hline 
  \end{tabular} \end{center}
    \caption{\sf \label{tab:7} The first three hadronic moments for the reference values of the input parameters and $E_{cut}=1$GeV, in the on-shell and kinetic schemes. 
  } 
\end{table}
As illustrated in Fig.~\ref{fig:hadfits} by a few examples, the precision of the fits to the functions 
$\chi_{0i}^{(2)}$ is similar to that of the functions $\eta_i^{(2)}$ entering the leptonic 
moments. On the other hand, the numerical evaluation of \cite{melnikov2} is 
very accurate in the case of $\chi_{ij}^{(2)}$ for $i>0$ and a quadratic dependence on 
$r$ has to be included in the functional form. As the cancellations between different 
contributions are not as severe as for leptonic moments,  
the accuracy of the non-BLM 
corrections is in fact a minor issue in the case of hadronic moments. 

\begin{table}
  \begin{center} \begin{tabular}{|c|lll|lll|}
    \hline 
     & \multicolumn{3}{|c|}{$\mu=1$GeV,\  $m_c^{\overline{\rm MS}}(2\rm GeV)$ }  &
        \multicolumn{3}{|c|}{$\mu=1$GeV,\  $m_c^{\overline{\rm MS}}(3\rm GeV)$ } \\
    \hline
    & $h_1$ & $h_2$ & $h_3$ & $h_1$ & $h_2$ & $h_3$\\ \hline
$1/m_b^3$ & 4.301& 0.551 &4.94  &4.020 &0.618 &5.02 \\
$O(\as)$ & 4.355&0.758  & 4.60 &4.192 &0.830 &4.79  \\
$O(\beta_0\as^2)$ &4.304 &0.936 &4.21 & 4.169 &1.015 & 4.49 \\
$O(\as^2)$ &4.328 &0.865(4) & 4.18(4) &4.245(1) & 0.922(5)& 4.38(4) \\
\hline
  \end{tabular} \end{center}
    \caption{\sf \label{tab:8} The first three hadronic moments for the reference values of the 
    input parameters and $E_{cut}=1$GeV, in the kinetic scheme with $\overline{\rm MS}$ charm 
    mass evaluated at $\mu=2$ and 3GeV, with $m_c(2{\rm GeV})=1.1$GeV and $m_c(3{\rm 
    GeV})=1$GeV. The uncertainty in the $O(\as^2)$ is larger in the second case because the $m_c/m_b$ value is closer to the edge of the range considered in \cite{melnikov2}. 
      } 
\end{table}
\begin{figure}
\begin{center}
\includegraphics[width=7cm]{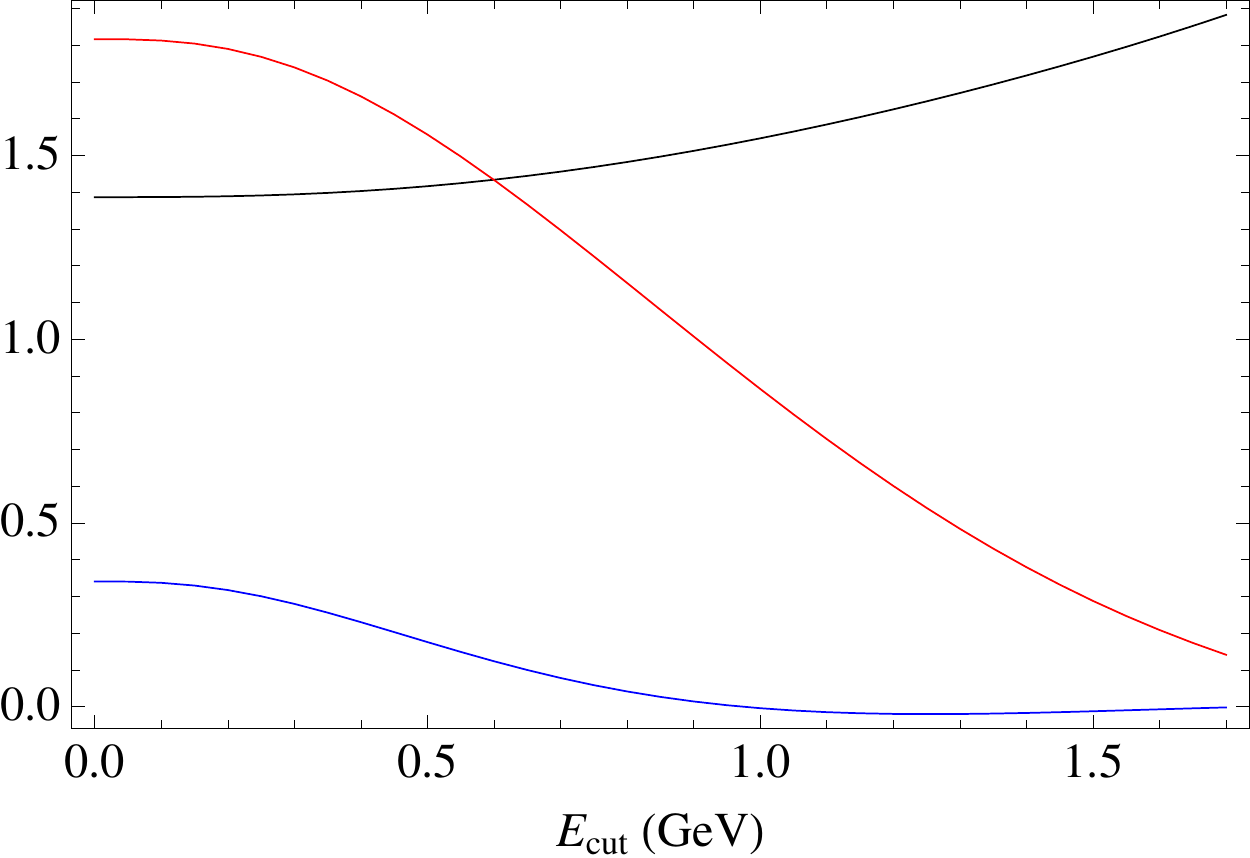} \ \ \
\includegraphics[width=7cm]{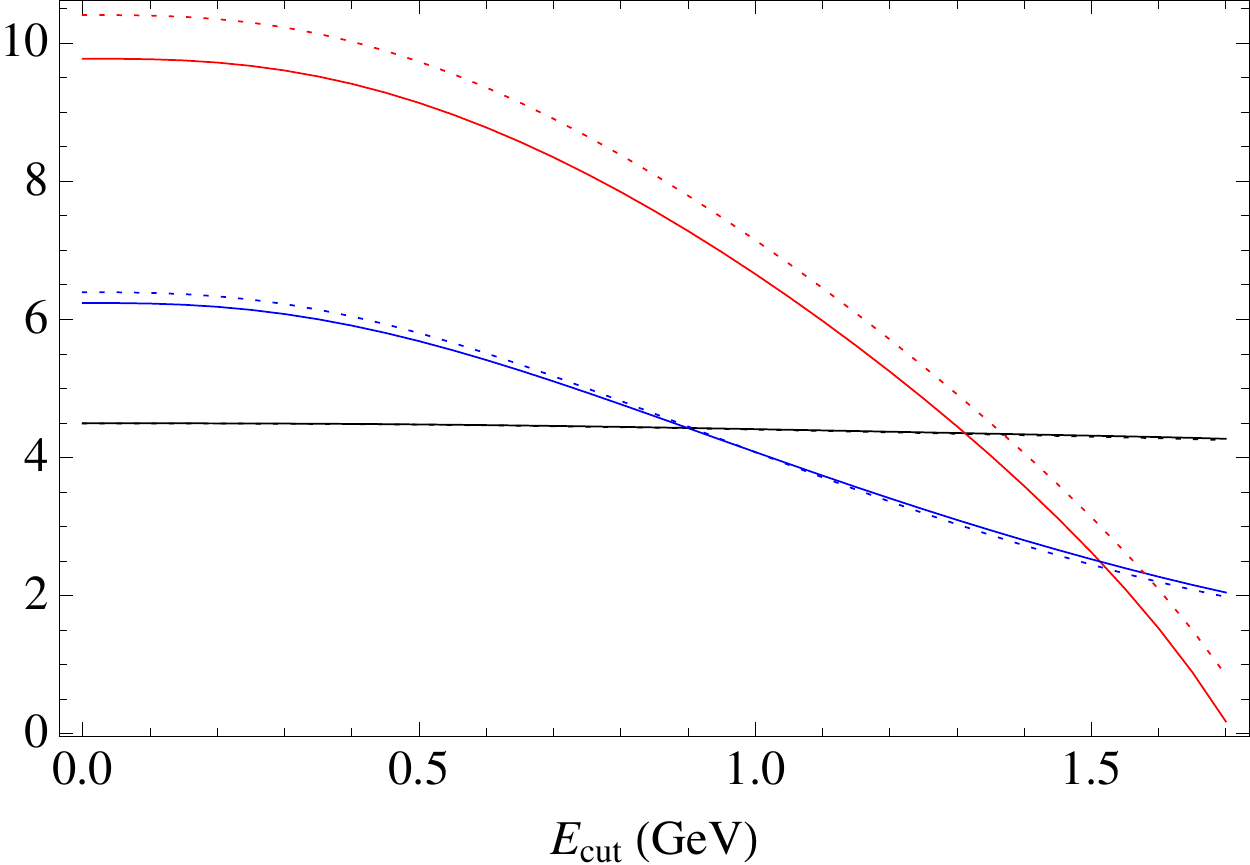} 
\caption{\sf $E_{cut}$ dependence of leptonic (left) and hadronic
 (right) at NNLO in the
kinetic scheme with $\mu=\mu_c=1 \GeV$. The black, red, 
 blue lines refer to $\ell_{1},10 \times\ell_2,-10\times\ell_3$  and to $h_1,8\times h_2,h_3$, 
 respectively, each expressed in \GeV\ to the appropriate power. The dotted lines 
 (indistinguishable for the leptonic moments) represent the predictions at $O(\as^2 \beta_0)$.}
\label{figxidep}
\end{center}
\end{figure}

Numerically, the non-BLM corrections in the on-shell scheme tend to partly compensate the
BLM corrections and, as perturbative corrections in general, are more important than in the
leptonic case, see Tables \ref{tab:6} and \ref{tab:7}.
The non-BLM effect is about $0.4\%$, 4\%, 3\%  for $h_{1,2,3}$, 
respectively.
In the kinetic scheme with $\mu=1\GeV$ the non-BLM corrections to the first and third 
moment are slightly suppressed, while the non-BLM correction to $h_{2}$ is larger than in 
the on-shell scheme (7-9\%), see also   Fig.~\ref{figxidep}.  It is worth reminding that 
the perturbative contributions to the building blocks $M_{11}$, $M_{20}$ and $M_{10}$ 
(all vanishing at tree-level) are dominant in $h_2$, and that these building blocks 
are known only at the next-to-leading order level. This explains the large relative 
value of the non-BLM corrections and implies a commensurate uncertainty. But the 
results shown in the Tables hide an important cancellation between the $\mupi$ and $\rd$
contributions to $h_2$ for the reference values of Eq.~(\ref{inputs}): for $E_{cut}=1\GeV$ they are about $+1.6$ and $-1.2\GeV^4$, 
respectively. 
Non-perturbative contributions indeed dominate the higher moments $h_{2,3}$, which are therefore subject to much larger uncertainties 
from both higher order terms in the OPE and the missing $O(\as)$ corrections to the Wilson 
coefficients of the dimension 5 and 6 operators.

\subsection{Scale dependence of hadronic moments}
\begin{figure}[t]
\begin{center}
\includegraphics[width=5.2cm]{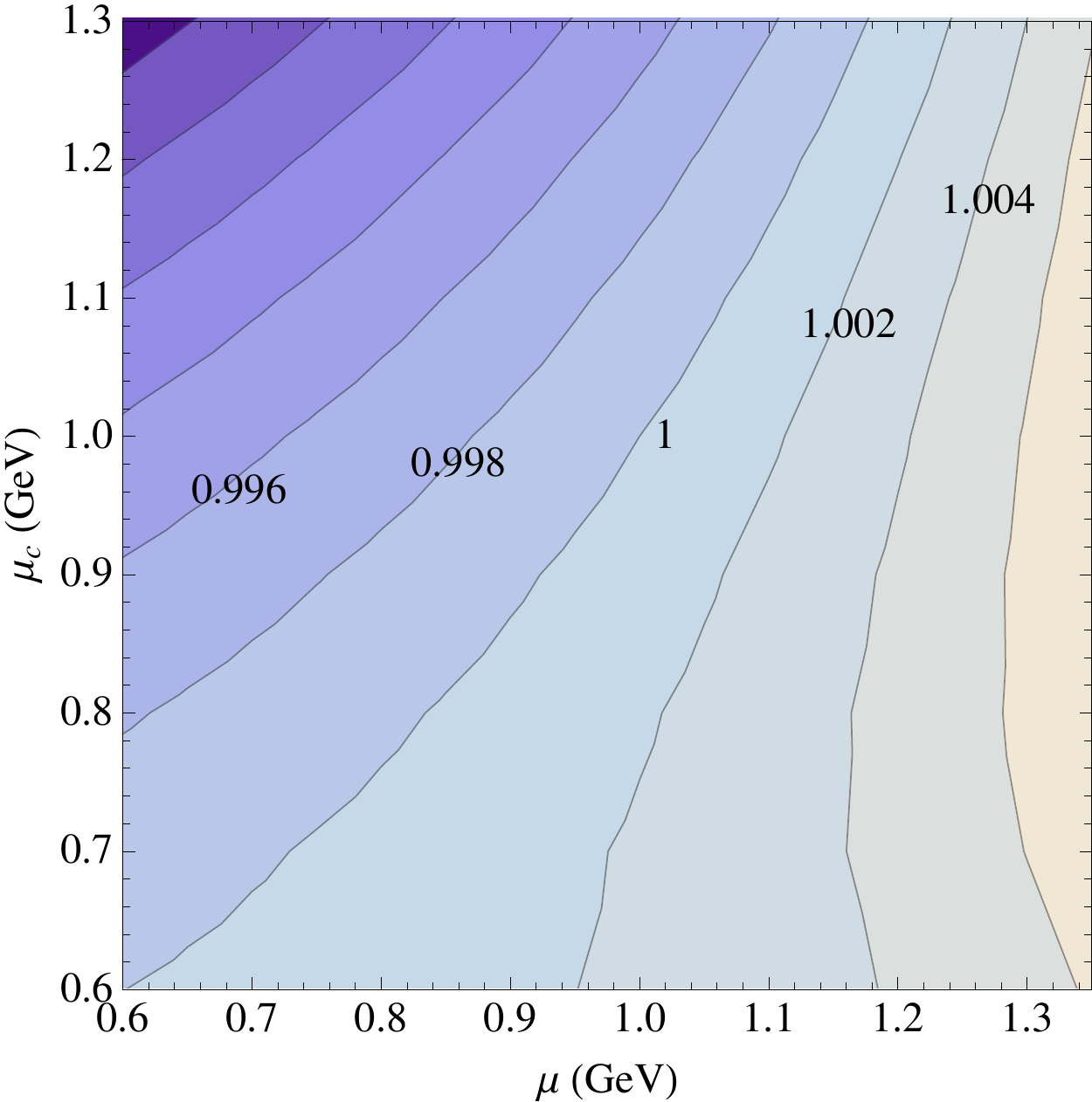}\ 
\includegraphics[width=5.2cm]{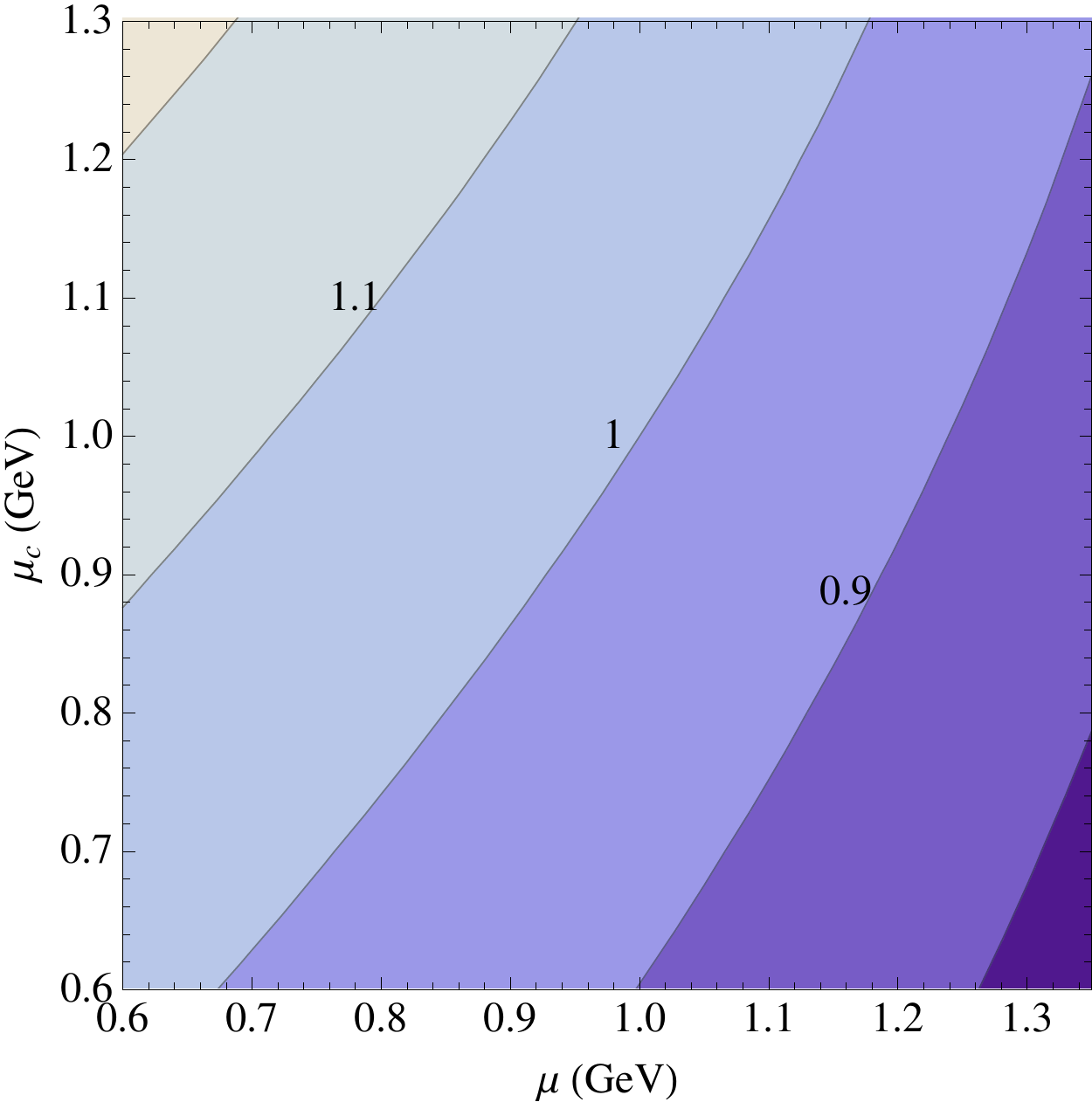}\
\includegraphics[width=5.2cm]{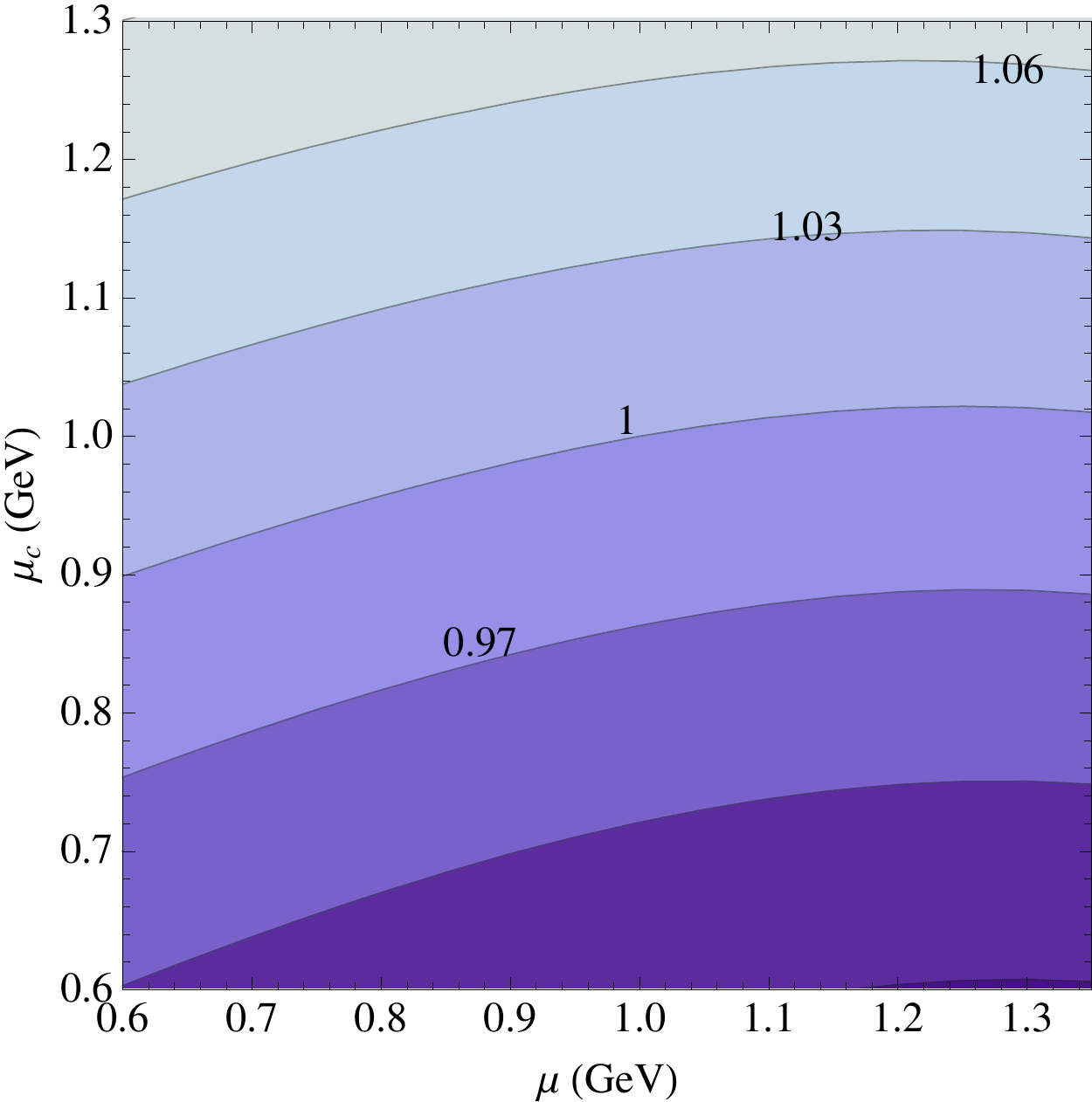}
\caption{\sf $\mu$ and $\mu_c$ dependence of the first three hadronic central moments at $E_{cut}
=1\GeV$ normalized to their reference value $\mu=\mu_c=1\GeV$. 
The three plots refer to $h_{1,2,3}$, respectively.  }
\label{figkinmu2}
\end{center}
\end{figure}

The $\mu$ and $\mu_c$ dependence of the hadronic moments in the kinetic scheme is shown in Fig.~\ref{figkinmu2} for $E_{cut}=1\GeV$.
The first moment has very small residual scale dependence, like the leptonic moments.
Since the second and third hadronic moments receive much larger
power and perturbative contributions, the larger $\mu$ and especially $\mu_c$ dependence
is not surprising. 
\begin{figure}[t]
\begin{center}
\includegraphics[width=7cm]{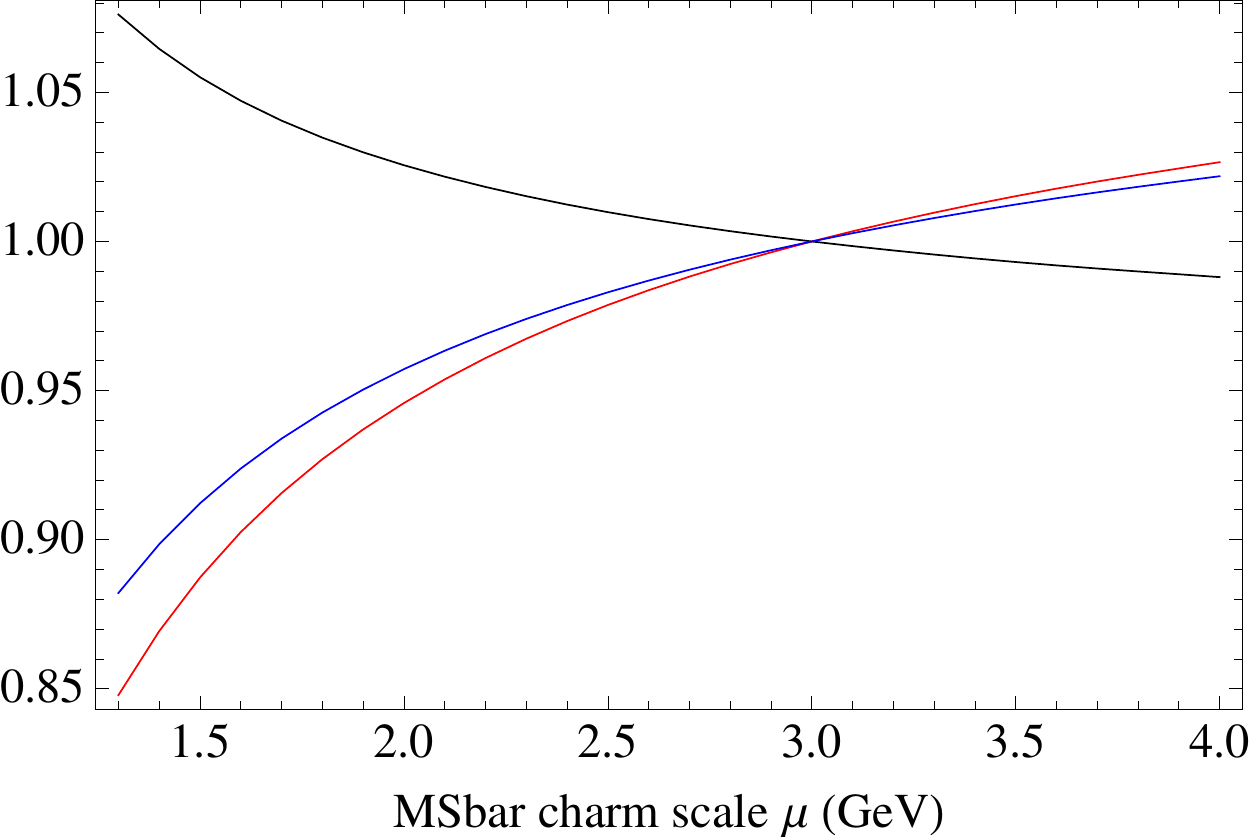} 
\includegraphics[width=7cm]{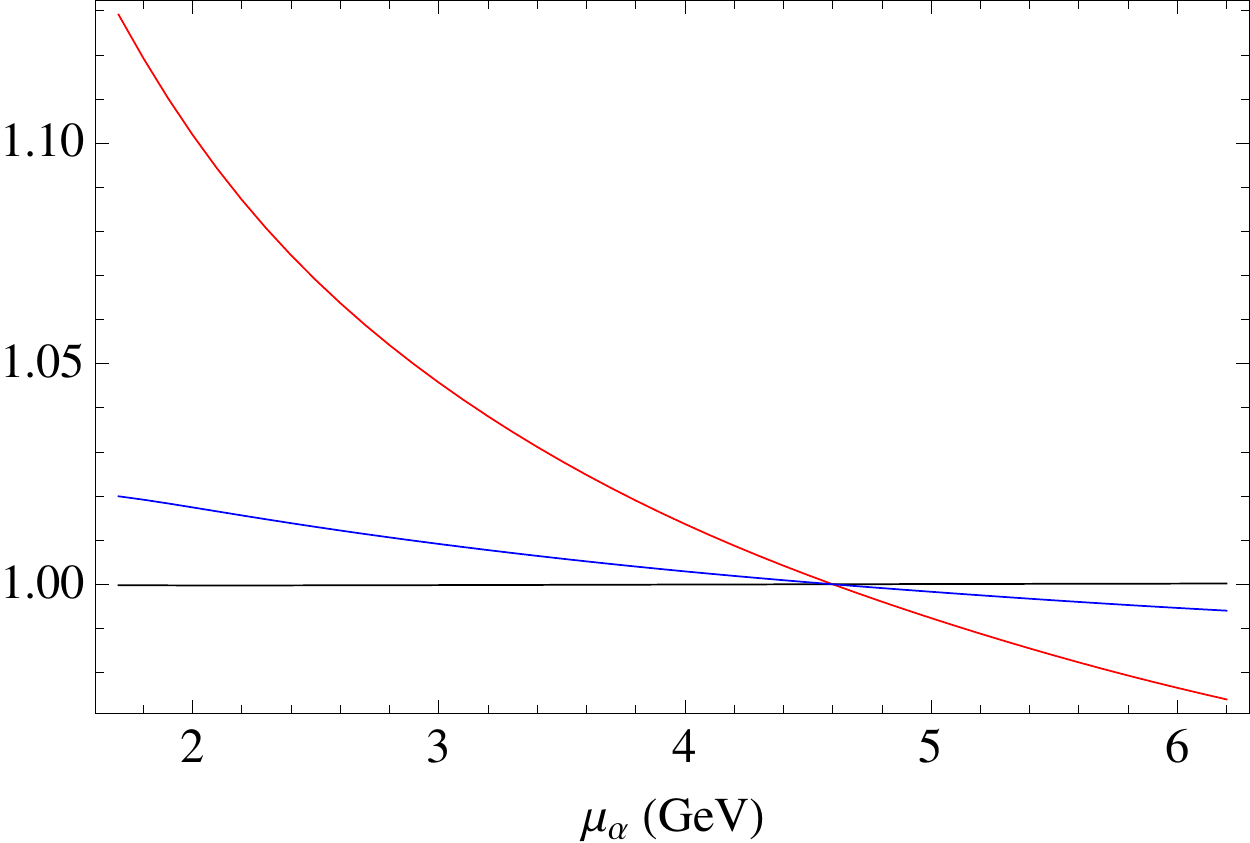} 
\caption{\sf Dependence of $h_{1,2,3}$ at $E_{cut}=0$ on 
$a)$ the charm $\overline{\rm MS}$ scale $\bar\mu$ (left panel) and $b)$  the scale of $\as$ 
when the kinetic scheme is applied to $m_c$ as well (right panel). The moments are 
normalized to their values at $\bar\mu=3\GeV$ and $\mu_\alpha=m_b$, respectively.
 The kinetic cutoff is fixed at 1 \GeV. The black, red, 
 blue lines refer to $h_{1,2,3}$.}
\label{figmsmu2}
\end{center}
\end{figure}
The results of the calculation of the moments when the charm mass is renormalized in the 
 $\overline{\rm MS}$ scheme are illustrated in Table \ref{tab:8} for two different values of 
 the charm scale $\bar\mu$. It is interesting that for $\bar\mu=3\GeV$ the change of scheme 
 induces large non-BLM corrections, suggesting that $\bar\mu=2\GeV$ might be a better choice.
 Fig.~\ref{figmsmu2} is the analogue of Fig.~\ref{figmsmu1} for the hadronic moments. 
 The plot on the left  shows the dependence on  the charm mass scale $\bar \mu$, 
 when $h_{1,2,3}$ are normalized to their values for $\bar\mu=3\GeV$, $m_c(3\GeV)
 =1\GeV$. In this case the $\bar\mu$ dependence is even larger than in the leptonic case,
 but again it should not be interpreted as indication of large higher order corrections in the 
 calculation of the moments and one should preferably use values $\bar\mu$ that minimize 
 the corrections to the moments.
 
Like the leptonic moments,
the first hadronic moment can be used to constrain the heavy quark masses. The higher 
hadronic moments are too sensitive to higher dimensional expectation values to play a role 
in this respect.  
At reference values of the inputs and for $E_{cut}=1\GeV$, a small shift in $m_{b,c}$ 
induces
\be\delta h_1= -4.84\, \delta m_b +3.08\, \delta m_c, 
\nonumber\ee
which vanish for $\delta m_c/\delta m_b\approx 1.6$. Therefore, $h_1$ gives a constraint
in the $(m_c,m_b)$ plane very similar to the leptonic central moments. The considerations 
on scheme dependence made after Eq.~(\ref{schemechange}) apply to the first hadronic 
moment as well. 

Finally, in the right-hand plot of Fig.~\ref{figmsmu2} we show the dependence of the kinetic 
scheme results on the $ \overline{\rm MS}$ scale of $\as$, $\mu_\alpha$, for fixed cutoff $
\mu=\mu_c=1\GeV$. The evolution of $\as$ is 
again computed at 4 loops, using  $\as(m_b)=0.22$ as input. Due to its direct connection 
with the size of the perturbative contributions,  the $\mu_\alpha$ dependence  is small for 
$h_1$. 
The strong residual dependence of $h_2$, 
on the other hand, suggests that in the absence of higher order corrections, it would be  
prudent to choose $\mu_\alpha$ close to $m_b/2$.  

The overall theoretical errors computed using the method proposed in
 \cite{btoc} are reported in the 
last row of Tables \ref{tab:6} and \ref{tab:7}: they are generally a factor two or more larger than the present experimental ones.
Cutting by half the variation in $m_{b,c}$ and $\as$, as it was proposed in the previous Section,  leads to a 25-30\% smaller total uncertainty on  $h_1$, while the error on $h_{2,3}$ is almost unaffected.
 The calculation of $O(\as)$ corrections to the Wilson coefficients and 
that of higher orders in the OPE \cite{Mannel:2010wj}, will have a stronger impact on the
total uncertainty. We also recall that the $O(1/m_b^{4,5})$ contributions to $h_i$ found in 
Ref.~\cite{Mannel:2010wj} are of the same order of magnitude of the total error in the last 
row of Tables \ref{tab:6} and \ref{tab:7}, while those of $O(\as\mu_\pi^2)$ \cite{Becher:2007tk}
are comparatively small.

\section{Mass scheme conversion}
Semileptonic moments provide interesting constraints on the bottom and charm masses, 
which can be compared and combined with other experimental determinations of these parameters. 
Since in many applications the quark masses are renormalized 
in the $\overline{\mbox{MS}}$, we will now  review the relation between the kinetic and $\overline{\mbox{MS}}$ 
definitions of the heavy quark masses.  Accurate conversion formulas can also be used 
to study the scheme  dependence of our results. 
 The perturbative relation 
between  the quark pole mass and the kinetic mass is known completely at $O(\as^2)$  and
$O(\as^3\beta_0^2)$    \cite{Czarnecki:1997sz}. Combining 
it with the one between pole and $\overline{\mbox{MS}}$ masses, known to $O(\as^3)$ \cite{mmsbar}, one gets
the following NNLO expansion for the ratio between the $\overline{\mbox{MS}}$ mass $\overline m(\bar\mu)$ and its kinetic counterpart $m(\mu)$
\begin{eqnarray}
\frac{\overline{m}(\bar{\mu})}{m(\mu)}&=&
1-\frac{4}{3}\frac{\as^{(n_l)}(M)}{\pi}\left(1-\frac{4}{3}\frac{\mu}{m(\mu)}
-\frac{\mu^2}{2m(\mu)^2}+\frac{3}{4} \ L\right) \nonumber\\
&& + \ \left(\frac{\as}{\pi}\right)^2
\Bigl\{ \Bigl[-\frac{2}{3} L+\left(\frac{1}{3} \ln
   \frac{M}{2 \mu}+\frac{13}{18}\right) \beta^{(n_l)} _0- \frac{\pi
   ^2}{3}+\frac{47}{18}\Bigr] \,\frac{\mu^2}{m^2} \nonumber\\
&& +\Bigl[-\frac{16}{9}
   L+\left(\frac{8}{9} \ln \frac{M}{2
   \mu}+\frac{64}{27}\right) \beta^{(n_l)} _0-\frac{8 \pi
   ^2}{9}+\frac{188}{27}\Bigr] \,\frac{\mu}{m} 
   -\frac43 \sum_i^{n_l} \Delta\left(\frac{m_i}{m}\right)
   \nonumber\\
&& +\frac{7}{12} L^2-\frac{2 L}{9}-\Bigl[\frac{L^2}{8} +\frac{13 L}{24}+\ln
   \frac{M^2}{\bar{\mu }^2} \left(\frac{L}{4} +\frac{1}{3}\right)+\frac{\pi ^2}{12}+\frac{71}{96}\Bigr]
   \beta^{(n_l)} _0 \nonumber\\
&& +\frac{\zeta (3)}{6}-\frac{\pi^2}{9} \ln 2+\frac{7 \pi
   ^2}{12}-\frac{169}{72} + \left(\frac{2}{9}+\frac{L}{6} \right)
   \ln\frac{m_h^2}{\bar{\mu }^2}\Bigr\} ,
\label{eq:ratioMbarMu2mkin}
\end{eqnarray}
where we have used
\be
L = \ln\left[\bar{\mu}^2/m(\mu)^2\right],\quad\quad\beta^{(n_l)}_0=11-2/3 \ n_l,
\quad \quad \Delta(x)=\frac{\pi^2}8 x -0.597 x^2 +0.230 x^3.
\nonumber\ee
The function $\Delta(x)$ arises from the two-loop diagrams with a closed massive quark 
loop in the relation between pole and $\overline{\mbox{MS}}$ mass \cite{gray,rundec}. It 
represents the correction to the massless loop  and  is only relevant in the case of charm 
mass effects on the $b$ mass.
 As mentioned in  footnote \ref{foot:1}, such effects are not included  in the 
 conversion to the kinetic scheme,  which is performed assuming charm to be 
 decoupled \cite{Czarnecki:1997sz}\footnote{Finite charm mass $O(\as^2)$ effects in the 
 transition between pole and kinetic mass can in principle 
 computed from Eq.~(11) of \cite{Czarnecki:1997sz}.}.
There is also a dependence on $m_h$, the mass scale that defines the threshold 
between $n_l$ and $n_l+1$ quark 
flavors, generally set equal to $m$. 
In the limit $\mu\to 0$ the above formula relates the pole and the $\overline{\mbox{MS}}$ masses.

When the scale of the $\overline{\mbox{MS}}$ mass is equal to the $\overline{\mbox{MS}}$ mass itself and the coupling constant  is evaluated at the mass scale the above formula reduces to
\begin{eqnarray}
 \frac{\overline{m}(\overline{m})}{m(\mu)}&=&
1-\frac{4}{3}\frac{\as(m)}{\pi}\left(1-\frac{4}{3}\frac{\mu}{m(\mu)}
-\frac{\mu^2}{2m(\mu)^2}\right) \nonumber\\
&& + \ \left(\frac{\as}{\pi}\right)^2
\Bigl\{
   \Bigl[\left(\frac{1}{3} \ln
   \frac{m}{2 \mu}+\frac{13}{18}\right) \beta^{(n_l)} _0-\frac{\pi
   ^2}{3}+\frac{23}{18}\Bigr] \frac{\mu^2}{m^2} \nonumber\\
&& +\Bigl[\left(\frac{8}{9} \ln \frac{m}{2
   \mu}+\frac{64}{27}\right) \beta^{(n_l)} _0-\frac{8 \pi
   ^2}{9}+\frac{92}{27}\Bigr]\, \frac{\mu}{m}  -\Bigl(
   \frac{\pi ^2}{12}+\frac{71}{96}\Bigr)
   \beta^{(n_l)} _0 
\nonumber\\
&&  +  \frac{\zeta (3)}{6}-\frac{\pi^2}{9} \ln 2
 +\frac{7 \pi
   ^2}{12}+\frac{23}{72} 
   -\frac43 \sum_i^{n_l} \Delta\left(\frac{m_i}{m}\right)
     \Bigr\} .
\label{eq:ratioMbarMbar2mkin}
\end{eqnarray}
This result differs from those reported in \cite{Czarnecki:1997sz,Benson:2003kp}
by a term $\left(\frac{\as}{\pi}\right)^2 \left(\frac{\pi^2}{18}+\frac{71}{144}\right)$. The difference can be traced 
back to a mismatch between the number of quarks considered in the $\overline{\mbox{MS}}$--pole relation. In the limit $\mu\to0$, Eq.~(\ref{eq:ratioMbarMbar2mkin}) reproduces
the standard two-loop $\overline{\mbox{MS}}$--pole relation, {\it cf.} Eq.~16 of \cite{rundec} and \cite{gray}. The  BLM corrections of $O(\as^3\beta_0^2)$ have been also considered 
in \cite{Czarnecki:1997sz,Benson:2003kp}.
\subsection{Charm mass conversion}
In the case of the charm mass we  employ the two above formulas with  $n_l=3$,
 neglecting all $\Delta$ functions.  Eq.~\eqref{eq:ratioMbarMbar2mkin} with
$\as(m_c)=0.41$ and $m_c(1\GeV)=1.15$ 
gives 
\bea
\overline{m_c}(\overline{m_c})&=&1.150 +0.108_{\as} +0.071_{\as^2}=1.329\GeV  \,, 
\label{eq:mcmc}
 \eea
 where the one and two-loop contributions are identified by the subscript. Unsurprisingly, the perturbative 
 series converges poorly. One easily verifies that  BLM corrections dominate the two-loop contribution and that  the result depends strongly on the  scale of $\as$.
The apparent convergence of the perturbative series can be drastically improved if one normalizes the 
charm mass at higher $\overline{\mbox{MS}}$ scales.
In fact, in the evolution of $\overline{m_c}(\bar\mu)$ to  low scales the effect of RGE 
resummation  becomes significant, but these contributions are  absent 
from the fixed order NNLO calculation of the moments from which 
the kinetic mass could be extracted. In order to convert the results of a fit to the 
semileptonic moments in the kinetic scheme into the $\overline{\mbox{MS}}$ scheme (or {\it vice versa})  it is therefore important to adopt a scale 
$\bar\mu$ above 2 GeV. To illustrate the point let us use $\bar\mu=3 \,\GeV$ and adopt the 
central value of \cite{masses1}, $\overline{m_c}(3\rm GeV)=0.986\GeV$. 
Inverting Eq.~(\ref{eq:ratioMbarMu2mkin}) and employing 
$\as^{(3)}(3\rm GeV)=0.247$ (which corresponds to $\as^{(4)}(m_b)=0.22$ for $m_h=1.5\GeV$), we obtain the  kinetic charm mass
\be
m_c^{kin}({\rm 1 GeV})=0.986 +0.083_{\as} +0.022_{\as^2}=1.091\GeV     
 \label{eq:mckin3}
\ee
where the final result depends little on the scale of $\as$ and on its exact value. The difference 
\be
 m_c(1\GeV)-\overline{m_c}(\rm 3GeV)=(0.105\pm 0.015)\GeV
 \label{eq:mcdiff3}\ee
depends of course on the input $\overline{m_c}(3\rm GeV)$: for  
values between 0.95 and 1.05 \GeV\ the central value moves from 0.096 to 0.121\GeV. 

Using a lower scale $\bar \mu=2$ GeV the difference between kinetic and $\overline{\mbox{MS}}$ is even smaller. Using 
$\overline{m_c}(2\rm GeV)=1.092$ GeV, obtained from the four-loop evolution of the result of \cite{masses1}, one finds
\be
m_c^{kin}({\rm 1 GeV})=1.092 +0.037_{\as} -0.013_{\as^2}=1.116\GeV     
\label{eq:mckin2}
\ee
with very small dependence on the scale of $\as$. The difference between 
the final results in Eqs.~(\ref{eq:mckin3}) and (\ref{eq:mckin2}) is mostly due to our use of 
four-loop RGE to connect the inputs at $\bar\mu=3$ and 2 \GeV.
Our estimate is
\be
 m_c(1\GeV)-\overline{m_c}(2\rm GeV)=(0.024\pm 0.010)\GeV,
 \label{mcdiff2}
 \ee 
whose central value varies between 0.017  and 0.032  \GeV\ for $\overline{m_c}(2\rm GeV)$
in the range 1.05 to 1.15 \GeV. Eqs.~(\ref{eq:mcdiff3},\ref{mcdiff2}) can be used 
to compare accurately the predictions for the semileptonic moments when the kinetic and $
\overline{\mbox{MS}}$ schemes are used for $m_c$. It turns out that the size of the scheme 
dependence found in this way is similar to the uncertainty due to scale dependence 
investigated in Secs. 2 and 3.

Once the relation between kinetic and  $\overline{\mbox{MS}}$ masses is known at large $
\bar\mu$ one can use the four-loop RGE together with Eq.~(\ref{eq:mckin3},\ref{eq:mckin2})
to compute $\overline{m_c}(\overline{m_c})$. Using
the more precise Eq.~(\ref{eq:mckin2}) we obtain
\be
m_c(1\GeV)-\overline{m_c}(\overline{m_c})= 0.16\pm 0.02\GeV.\nn
\ee

\subsection{Bottom mass conversion}

In the case of the relation between $\overline{\mbox{MS}}$ and pole bottom mass 
the  effects related to the charm mass in closed quark loops and described by 
$\Delta(r)$    are small but not negligible. For realistic  values
of  the charm to bottom mass ratio $\Delta(r)$ compensates up to  40\% of the
light quark contribution, {\it i.e.}\ the term multiplied by $n_l$ in the fourth line of Eq.~(\ref
{eq:ratioMbarMu2mkin}). Since the corresponding effects are not included in the definition of 
kinetic $m_b$,
in applying Eqs.~(\ref{eq:ratioMbarMu2mkin},\ref{eq:ratioMbarMbar2mkin}) to the 
bottom mass we therefore have the following  options
\begin{itemize}
\item [$(a)$] we decouple charm completely and set $n_l=3$, $\Delta(r)=0$;
\item [$(b)$] we treat charm as massless and set $n_l=4$, $\Delta(r)=0$;
\item [$(c)$] we decouple charm in the kinetic part of the calculation only, set $n_l=3$,  and replace $\Delta(r)
\to \Delta(r)- \frac{\pi^2}{24}-\frac{71}{192}$.
\end{itemize}
It is clear that the correct result lies somewhere between $(a)$ and $(b)$. On the other 
hand, since the NNLO expressions of the semileptonic moments have been derived 
under the assumption of charm decoupling in the relation between pole and kinetic masses,
but we have otherwise kept the charm mass effects, 
option $(c)$ is the most appropriate to convert 
the kinetic mass extracted from a NNLO fit to semileptonic moments into 
the $\overline{\mbox{MS}}$ scheme.
\begin{figure}[t]
\begin{center}
\includegraphics[width=8cm]{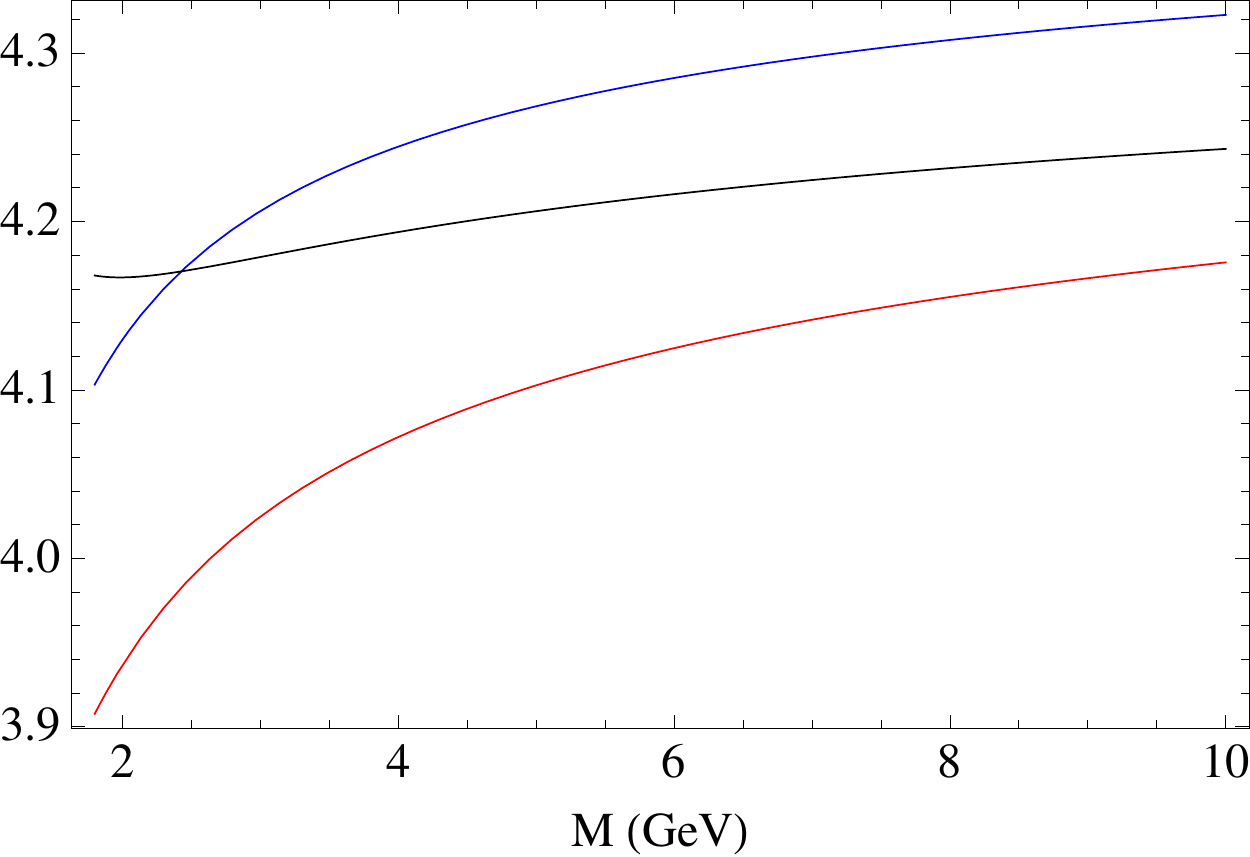}
\caption{\sf $M$ dependence of $m_b(m_b)$ in case $(c)$ for $m_b(1\GeV)=4.55\GeV$. The blue, red, black curves refer to one-loop, BLM, two-loop, respectively.}
\label{fig:mbmbscale}
\end{center}
\end{figure}

Let us now illustrate the three options computing the $\overline{\mbox{MS}}$ bottom mass 
from the kinetic one at a reference value $\mu=1\GeV$.  
We use $m_b(1\GeV)=4.55\GeV$ and $\as(m_b)=0.22$
\bea
\overline{m_b}(\overline{m_b})&=&4.550 -0.290_{\as} -0.073_{\as^2}=4.187\GeV,     \quad\quad (a)\nn\\
&& 4.550-0.290_{\as} -0.060_{\as^2}=4.200\GeV,   \quad\quad (b)\label{eq:mbmb}\\ 
&& 4.550-0.290_{\as} -0.057_{\as^2}=4.202\GeV,   \quad\quad (c)\nn
\eea
from which we see that option $(c)$ is numerically very close to $(b)$.
\begin{table}
  \begin{center} \begin{tabular}{|c|lll|}
    \hline 
    &&$\as(m_b)$& \\ \hline
   $m_b(1\GeV)$ 
    &0.21&$0.22$ &$0.23$ \\ \hline
  4.500   & 4.150& 4.127 &4.103\\
 4.550  & 4.194 &4.171  &4.147\\
 4.600 & 4.239 & 4.216 & 4.192 \\
    \hline 
  \end{tabular} \end{center}
    \caption{\sf \label{tab:mbmb} $\overline{m_b}(\overline{m_b})$ for different values of the kinetic mass and of $\as(m_b)$ in  case $(c)$ with $M=2.5\GeV$. } 
\end{table}
The BLM corrections are large in 
Eq.~(\ref{eq:mbmb}) but partly (50\%) compensated by sizable non-BLM ones. 
The $O(\as^3\beta_0^2)$ corrections \cite{Czarnecki:1997sz} shift 
$\overline{m_b}(\overline{m_b})$ further down by about 20 MeV.
The dependence of the NNLO  result on the scale of $\as$ is shown in Fig.~\ref{fig:mbmbscale}: 
it suggests a residual uncertainty of  about $40$ MeV.  Table \ref{tab:mbmb} shows the 
value of $\overline{m_b}(\overline{m_b})$ for different values of the inputs, using for $\as$ a 
lower scale than $m_b$ itself, $M=2.5\GeV$. The dependence on $r$ is negligible. 

If one computes $\overline{m_b}$ from $m_b(1\GeV)$ at a scale around 3 GeV the 
perturbative corrections are smaller because the numerical values of the masses get closer.
For instance one gets  $\overline{m_b}(2.8\GeV)=4.550+0.023_{\as}-0.024_{\as^2}=4.550$.
This result can be evolved up using four-loop RGE, finding 
$\overline{m_b}(\overline{m_b})=4.182$,  which is consistent with the values in Table \ref{tab:mbmb}
but has arguably a slightly smaller error.
Our final estimate is
$$ m_b(1\GeV)-\overline{m_b}(\overline{m_b})=(0.37\pm 0.03)\GeV,$$
The error also includes in quadrature the parametric $\as$ error, $\delta \as(m_b)=0.007$.

\section{Summary}
I have presented the results of a complete $O(\as^2)$ calculation of the moments 
of the lepton energy and hadronic invariant mass distributions in semileptonic $B\to X_c \ell
\nu$ decays. The numerical Fortran code\footnote{The Fortran code is available from \tt gambino@to.infn.it} that has been  developed allows us to work in any perturbative scheme;  
I have adopted  the kinetic scheme, but have also considered the option of 
the $\overline{\rm MS}$ scheme for the definition of the charm mass.  
The non-BLM corrections are generally small and within the expected range. 
The new code will  allow for a NNLO fit to the semileptonic moments, 
with the possible inclusion of precise mass constraints, in a variety of perturbative schemes.
To facilitate the inclusion of mass constraints obtained in the $\overline{\rm MS}$ scheme, 
 I have critically reexamined the conversion from that scheme to the kinetic scheme for the heavy quark masses. Additional details 
concerning the NNLO prediction of the total rate can be found in the Appendix.

I  have performed a detailed study of the 
dependence of the results on various unphysical scales, obtaining useful indications on the 
size of the residual perturbative uncertainty in the moments, 
which is now definitely  smaller than the 
uncertainty due to higher dimensional contributions in the OPE and to $O(\as)$ contributions
to the Wilson coefficients of dimension 5 and 6 operators. 
Fortunately, work on both these 
aspects is progressing  and 
there are good prospects for a more accurate prediction of the inclusive semileptonic 
moments and a   more precise determination of $|V_{cb}|$.

\section*{Acknowledgements}
I am  grateful to Paolo Giordano for his contribution in the early stage of this work and to 
Daniele Papalia for carefully  checking many numerical results.  
I also thank Kirill Melnikov, Alexei Pak, and Einan Gardi
for useful communications and discussions, and Kolya Uraltsev for many clarifying conversations and for carefully reading a preliminary version of this paper. Work partly supported by MIUR under contract  2008H8F9RA$\_$002.

\section*{Appendix}
The determination of $|V_{cb}|$ from inclusive semileptonic 
$B$ decays is based on the expression for the charmed total semileptonic width. In this Appendix we update the numerical analysis in the kinetic scheme given in Ref.~\cite{Benson:2003kp} with the inclusion 
of the exact two-loop correction computed in \cite{melnikov,czarnecki-pak,melnikov2}. 
Keeping only terms through $O(1/m_b^3)$, the general structure of the width is (see \cite{Benson:2003kp} and refs.\ therein) 
 \bea 
\label{crate}
\Gamma[\bar{B} \to X_c e \bar{\nu}] &=& \frac{G_F^2 \,m_b^5}{192\pi^3} |V_{cb}|^2
g(\rho)  \left(1+A_{ew}\right) \left[ 1 +  \frac{\alpha_s(m_b)}{\pi} p_c^{(1)}(\rho,\mu) 
              + \frac{\alpha_s^2}{\pi^2} p_c^{(2)}(\rho,\mu) 
\right. \nonumber\\ && \hspace{1cm} \left.            
  -\frac{\mupi}{2 m_b^2}               +         \left( \frac{1}{2} - \frac{2 (1-\rho)^4}{g(\rho)} \right) \frac{\mug-\frac{\rls+\rd}{m_b}}{m_b^2} 
           +  \frac{d(\rho)}{g(\rho)}              \frac{\rd}{m_b^3}                           \right],
\eea
where  $\rho =( m_c/m_b)^2$, $g(\rho) = 1 - 8 \rho + 8 \rho^3 - \rho^4 - 12 \rho^2 \ln \rho$, 
$d(\rho)= 8 \ln \rho   -\frac{10 \rho^4}{3}+\frac{32 \rho^3}{3}-8 \rho^2-\frac{32 \rho}{3}
            +\frac{34}{3}$,
and all the masses and OPE parameters are defined
in the kinetic scheme with cutoff $\mu$. 
The expression for the width is  $\mu$-independent through $O(\as^2)$. 
 The term $A_{ew}=2 \alpha/\pi \ln M_Z/m_b\simeq
0.014$ is due to the electromagnetic running of the four-fermion operator from the weak to the  $b$ scale and represents the leading electroweak correction \cite{sirlin}.

The results of the two-loop BLM and non-BLM perturbative corrections  
\cite{Aquila:2005hq,czarnecki-pak,melnikov2} in the on-shell scheme are well 
approximated in the relevant mass range
by a simple interpolation formula, valid for $0.17<r<0.3$, 
\bea
p_c^{(2)}(r,0) &=& (-3.381 + 7.15\, r - 5.18\, r^2)\,
\beta_0   + 7.51 - 21.46\, r + 19.8\, r^2.\label{pc2}\nonumber
\eea
Applying these results and the well-known one-loop contribution \cite{nir} to the kinetic 
scheme calculation with $\mu=1\GeV$,  $m_b=4.6\GeV$, and $r=0.25$, the NNLO 
perturbative  series becomes
\be
\Gamma[\bar{B} \to X_c e \bar{\nu}] \propto 1 - 0.96\, \frac{\as}{\pi} -0.48 \,\beta_0 \left( \frac{\as}{\pi} \right)^2 + 0.81 \left( \frac{\as}{\pi} \right)^2  \approx 0.916
\label{expkin}
\ee
where $\beta_0=9$ corresponds to three light flavors, and we have used $\as(m_b)=0.22$
for the numerical evaluation. 
For comparison, Ref.~\cite{Benson:2003kp} used the same inputs and had $-1.0$  as 
coefficient of $\as^2$ and the  global perturbative factor of Eq.~(\ref{expkin}) was 0.908.
Of course, the difference is due to the recent complete NNLO calculation of \cite{melnikov,czarnecki-pak}.
As already discussed, finite charm 
quark mass effects in the NNLO conversion to the kinetic scheme have not been included. They can be expected to decrease the above value by up to 
0.002. Higher order BLM corrections are also known  \cite{pc2blm}
and have been studied in the kinetic scheme where  
the resummed BLM result is numerically very close to the NNLO one \cite{Benson:2003kp}. 
The residual 
dependence of the overall perturbative factor on the scale of $\as$ is mild: changing it 
between $m_b$ and $m_b/2$ decreases by less than 1\%, which represents a
reasonable estimate for the residual perturbative uncertainty.

We also provide an approximate formula for $|V_{cb}|$ in terms of the OPE parameters
and  $\as$.
Using $\tau_B=1.582$ ps and BR$_{sl,c}=0.105$ (the measured semileptonic branching ratio after subtraction of $b\to u l \nu$ transitions) we have
\bea
\frac{|V_{cb}|}{0.0416}&=&1+0.275\,(\as-0.22)-0.65\, (m_b-4.6)+0.39\,(m_c-1.15)+0.014 \,(\mu_\pi^2-0.4) \label{unfactor}
\nonumber\\&&+0.055\,(\mu_G^2-0.35)+0.10\, (\rho_D^3-0.2)-0.012\, (\rho_{LS}^3+0.15),
\eea
where all dimensionful quantities are expressed in GeV to the appropriate power and are understood in the kinetic scheme with $\mu=1\GeV$.
The numerical accuracy of this formula is limited to 1\% in the $1\sigma$ range of the 
current HFAG fits \cite{HFAG}. 
The combination of $m_{c,b}$ in the  above expression is very similar to the one that appears in the leptonic moments, see Eq.~(\ref{schemechange}).

If the $\overline{\rm MS}$ scheme is adopted for the charm mass with 
 $\overline m_c(2\GeV)=1.12\GeV$ and $m_b(1\GeV)=4.6\GeV$,  the NNLO perturbative series expressed in terms of $\alpha_s(m_b)$ becomes
\be
\Gamma[\bar{B} \to X_c e \bar{\nu}] \propto 1 - 1.27\, \frac{\as}{\pi} -0.33 \beta_0 \left( \frac{\as}{\pi} \right)^2 - 0.29 \left( \frac{\as}{\pi} \right)^2  \approx 0.895.
\label{pfactMS}
\ee
A simple test of the residual scheme dependence consists in comparing the products of the function $g(\rho)$ with  the overall perturbative factor,   Eqs.~(\ref{expkin},\ref{pfactMS}), 
computed in different schemes for $m_c$.  This product should be scheme independent when the same bottom mass is employed. In the case considered here the values 
$m_c(1\GeV)=1.15\GeV$ and $\overline m_c(2\GeV)=1.12\GeV$ satisfy Eq.(\ref{mcdiff2})
and the product is scheme independent to excellent approximation. 
For a larger $\overline{\rm MS}$ scale $\bar\mu= 3\GeV$ the perturbative expansion 
converges slower, but the scheme dependence is well within the 1\% uncertainty estimate.


\begin{thebibliography}{999}


\bibitem{Bigi:1992su}
  I.~I.~Y.~Bigi, N.~G.~Uraltsev and A.~I.~Vainshtein,
  Phys.\ Lett.\ B {\bf 293} (1992) 430 [Erratum-ibid.\  B {\bf 297} (1993) 477]
  [arXiv:hep-ph/9207214];
  I.~I.~Y.~Bigi, M.~A.~Shifman, N.~G.~Uraltsev and A.~I.~Vainshtein,
  Phys.\ Rev.\ Lett.\ {\bf 71} (1993) 496 [arXiv:hep-ph/9304225].

\bibitem{Blok:1993va}
  B.~Blok, L.~Koyrakh, M.~A.~Shifman and A.~I.~Vainshtein,
  Phys.\ Rev.\  D {\bf 49} (1994) 3356 [Erratum-ibid.\ D {\bf 50} (1994) 3572]
  [arXiv:hep-ph/9307247];
  A.~V.~Manohar and M.~B.~Wise, Phys.\ Rev.\ D {\bf 49} (1994) 1310 [arXiv:hep-ph/9308246].

\bibitem{HFAG}
  D.~Asner {\it et al.} [Heavy Flavor Averaging Group Collaboration],
  [arXiv:1010.1589 [hep-ex]], see also {\tt http://www.slac.stanford.edu/xorg/hfag/} .

\bibitem{kinetic}
I.~I.~Y.~Bigi, M.~A.~Shifman, N.~Uraltsev and A.~I.~Vainshtein,
Phys.\ Rev.\ D {\bf 56} (1997) 4017
[arXiv:hep-ph/9704245]
and 
Phys.\ Rev.\ D {\bf 52} (1995) 196
[arXiv:hep-ph/9405410].


\bibitem{Benson:2003kp}
  D.~Benson, I.~I.~Bigi, T.~Mannel and N.~Uraltsev,
  Nucl.\ Phys.\  B {\bf 665}, 367 (2003)
  [hep-ph/0302262].


\bibitem{btoc}
  P.~Gambino and N.~Uraltsev,
  Eur.\ Phys.\ J.\  C {\bf 34}, 181 (2004)
  [hep-ph/0401063].

\bibitem{Benson:2004sg}
  D.~Benson, I.~I.~Bigi and N.~Uraltsev,
  Nucl.\ Phys.\  B {\bf 710}, 371 (2005)
  [hep-ph/0410080].

\bibitem{BF}
  O.~Buchmuller, H.~Flacher,
  Phys.\ Rev.\  {\bf D73 } (2006)  073008.
  [hep-ph/0507253].


\bibitem{Bauer:2004ve}
  C.~W.~Bauer, Z.~Ligeti, M.~Luke, A.~V.~Manohar and M.~Trott,
  Phys.\ Rev.\  D {\bf 70}, 094017 (2004)
  [hep-ph/0408002].

\bibitem{Aquila:2005hq}
  V.~Aquila, P.~Gambino, G.~Ridolfi and N.~Uraltsev,
  Nucl.\ Phys.\  B {\bf 719} (2005) 77
  [arXiv:hep-ph/0503083].

\bibitem{1mb3}
M.~Gremm and A.~Kapustin,  {\it Phys.\ Rev.}\ {\bf D55} (1997) 6924.

\bibitem{ckm10}
  P.~Gambino, C.~Schwanda,
    [arXiv:1102.0210 [hep-ex]].

\bibitem{Bigi:2001ys}
  I.~I.~Y.~Bigi, N.~Uraltsev,
  Int.\ J.\ Mod.\ Phys.\  {\bf A16 } (2001)  5201-5248.
  [hep-ph/0106346].

\bibitem{melnikov}
  K.~Melnikov,
  Phys.\ Lett.\  B {\bf 666} (2008) 336
  [arXiv:0803.0951 [hep-ph]].
  
  \bibitem{czarnecki-pak}
  A.~Pak and A.~Czarnecki,
  Phys.\ Rev.\ Lett.\  {\bf 100} (2008) 241807
  [arXiv:0803.0960 [hep-ph]];  Phys.\ Rev.\  {\bf D78 } (2008)  114015.
  [arXiv:0808.3509 [hep-ph]].

  \bibitem{melnikov2}
 S.~Biswas and K.~Melnikov,
  JHEP {\bf 1002} (2010) 089
  [arXiv:0911.4142 [hep-ph]].

\bibitem{Mannel:2010wj}
  T.~Mannel, S.~Turczyk and N.~Uraltsev,
  JHEP {\bf 1011}, 109 (2010)
  [arXiv:1009.4622 [hep-ph]].
  

\bibitem{Becher:2007tk}
  T.~Becher, H.~Boos and E.~Lunghi,
  JHEP {\bf 0712}, 062 (2007)
  [arXiv:0708.0855 [hep-ph]].

\bibitem{ewerth}
 T.~Ewerth, P.~Gambino and S.~Nandi,
  Nucl.\ Phys.\  B {\bf 830} (2010) 278
  [arXiv:0911.2175 [hep-ph]].

\bibitem{voloshin} M.~B.~Voloshin, Phys.\ Rev.\ {\bf D51} (1995) 4934.

\bibitem{vub}
P.~Gambino, P.~Giordano, G.~Ossola and N.~Uraltsev,
  JHEP {\bf 0710} (2007) 058
  [arXiv:0707.2493 [hep-ph]];
  M.~Antonelli {\it et al.},
  Phys.\ Rept.\  {\bf 494 } (2010)  197-414.
  [arXiv:0907.5386 [hep-ph]].

\bibitem{gg}
  P.~Gambino, P.~Giordano,
  Phys.\ Lett.\  {\bf B669 } (2008)  69-73.
  [arXiv:0805.0271 [hep-ph]].
  
  \bibitem{masses1}
  K.~G.~Chetyrkin 
  {\it et al.},
  Phys.\ Rev.\  {\bf D80 } (2009)  074010
  [arXiv:0907.2110 [hep-ph]] and  arXiv:1010.6157 [hep-ph].

    \bibitem{hoang}
     B.~Dehnadi, A.~H.~Hoang, V.~Mateu, S.~M.~Zebarjad,
  [arXiv:1102.2264 [hep-ph]].
    
  \bibitem{masseslat} 
  C.~McNeile {\it et al.} [HPQCD Collaboration],
  Phys.\ Rev.\  D {\bf 82} (2010) 034512
  [arXiv:1004.4285 [hep-lat]].

\bibitem{paz} G.~Paz,  arXiv:1011.4953 [hep-ph].
  
  
  \bibitem{jez} M.~Jezabek and J.~H.~Kuhn,
Nucl.\ Phys.\ B {\bf 314} (1989) 1;
Nucl.\ Phys.\ B {\bf 320} (1989) 20;
A.~Czarnecki, M.~Jezabek and J.~H.~Kuhn,
Acta Phys.\ Polon.\ B {\bf 20} (1989) 961;
A.~Czarnecki and M.~Jezabek,
Nucl.\ Phys.\ B {\bf 427} (1994) 3
[arXiv:hep-ph/9402326].

\bibitem{gremm}
M.~Gremm and I.~Stewart,
{\it Phys.\ Rev.}\ {\bf D55} (1997) 1226.

\bibitem{Czarnecki:1997sz}
  A.~Czarnecki, K.~Melnikov and N.~Uraltsev,
  Phys.\ Rev.\ Lett.\  {\bf 80} (1998) 3189
  [arXiv:hep-ph/9708372].
  
  
 \bibitem{rundec}
  K.~G.~Chetyrkin, J.~H.~Kuhn, M.~Steinhauser,
  Comput.\ Phys.\ Commun.\  {\bf 133 } (2000)  43-65.
  [hep-ph/0004189].
  
  \bibitem{Falk:1997jq}
  A.~F.~Falk, M.~E.~Luke,
  Phys.\ Rev.\  {\bf D57 } (1998)  424-430.
  [hep-ph/9708327];
  A.~F.~Falk, M.~E.~Luke, M.~J.~Savage,
  Phys.\ Rev.\  {\bf D53 } (1996)  2491-2505
  [hep-ph/9507284].
  
\bibitem{trott}
  M.~Trott,
  Phys.\ Rev.\  D {\bf 70} (2004) 073003
  [arXiv:hep-ph/0402120].

\bibitem{kolyapert}
  N.~Uraltsev,
  Int.\ J.\ Mod.\ Phys.\  A {\bf 20} (2005) 2099
  [arXiv:hep-ph/0403166].

\bibitem{melnikov-private} K.~Melnikov, private communication.

\bibitem{mmsbar}
K.~G.~Chetyrkin, M.~Steinhauser,
  Nucl.\ Phys.\  {\bf B573 } (2000)  617-651.
  [hep-ph/9911434];
K.~Melnikov, T.~v.~Ritbergen,
  Phys.\ Lett.\  {\bf B482 } (2000)  99-108
  [hep-ph/9912391].

\bibitem{gray}
N.~Gray, D.~J.~Broadhurst, W.~Grafe, K.~Schilcher,
  Z.\ Phys.\  {\bf C48 } (1990)  673-680.
  

\bibitem{sirlin}
  A.~Sirlin,
  Nucl.\ Phys.\  B {\bf 196} (1982) 83.

\bibitem{pc2blm}
 P.~Ball, M.~Beneke and V.~M.~Braun,
 Phys.\ Rev.\ D {\bf 52} (1995) 3929 [arXiv:hep-ph/9503492];
  see also   N.~Uraltsev,
  Nucl.\ Phys.\  {\bf B491 } (1997)  303-322 
  [hep-ph/9610425] and Ref.~\cite{Benson:2003kp}.

\bibitem{nir} Y.~Nir,
  Phys.\ Lett.\  B {\bf 221} (1989) 184.

\end{thebibliography}
\end{document}